\documentclass[prd,superscriptaddress,preprintnumbers,amsmath,amssymb,floatfix,nofootinbib]{revtex4}

\usepackage{graphicx}
\usepackage{amsmath}    
\usepackage[normalem]{ulem}
\usepackage{bm}

\usepackage[utf8x]{inputenc}
\usepackage{multirow}
\usepackage{subfigure}
\usepackage{booktabs}
\usepackage{epsfig,graphics}
\usepackage{amsfonts,amssymb,amsmath}            
\usepackage{amsthm}

\newcommand{\nc}{\newcommand}

\nc{\be}{\begin{equation}}

\nc{\ee}{\end{equation}}

\nc{\bea}{\begin{eqnarray}}

\nc{\eea}{\end{eqnarray}}

\nc{\xx}{\nonumber\\}

\nc{\ct}{\cite}

\nc{\la}{\label}

\nc{\eq}[1]{(\ref{#1})}

\theoremstyle{plain}

\theoremstyle{definition}

\begin{document}
\title{Test of Emergent Gravity}

\author{Sunggeun Lee}
\affiliation{Department of Physics, Sogang University, Seoul 121-741, Korea}

\author{Raju Roychowdhury}
\affiliation{Center for Quantum Spacetime, Sogang University, Seoul 121-741, Korea}
\affiliation{Center for Theoretical Physics and Department of Physics \& Astronomy,
Seoul National University, Seoul 151-747, Korea}

\author{Hyun Seok Yang}
\affiliation{Center for Quantum Spacetime, Sogang University, Seoul 121-741, Korea}

\date{\today}

\begin{abstract}

In this paper we examine a small but detailed test of the emergent gravity picture with explicit
solutions in gravity and gauge theory. We first derive symplectic $U(1)$ gauge fields starting
from the Eguchi-Hanson metric in four-dimensional Euclidean gravity. The result precisely
reproduces the $U(1)$ gauge fields of the Nekrasov-Schwarz instanton previously derived from
the top-down approach. In order to clarify the role of noncommutative spacetime,
we take the Braden-Nekrasov $U(1)$ instanton defined in ordinary commutative spacetime
and derive a corresponding gravitational metric. We show that the K\"ahler manifold determined
by the Braden-Nekrasov instanton exhibits a spacetime singularity
while the Nekrasov-Schwarz instanton gives rise to a regular geometry-the Eguchi-Hanson space.
This result implies that the noncommutativity of spacetime plays an important role
for the resolution of spacetime singularities in general relativity.
We also discuss how the topological invariants associated with noncommutative $U(1)$ instantons
are related to those of emergent four-dimensional Riemannian manifolds
according to the emergent gravity picture.

\end{abstract}


\maketitle

\section{Introduction}

In order to understand our physical world, it is necessary to take quantum mechanics to be superordinate
to classical mechanics. The famous two-slit experiment in quantum mechanics, for example, cannot be
explained by simply extrapolating classical physics to the atomic world. Rather classical physics must be
understood as phenomena emergent from quantum world when a certain limit is taken to a classical regime.
However, in formulating quantum mechanics, we often start by working in a purely classical language
that overlays quantum concepts upon the classical framework, that places quantum mechanics in a somewhat
secondary position. Fortunately, the strategy of beginning with a theoretical description
that is classical and then subsequently including the features of quantum mechanics has been extremely fruitful for many years though it may be too conservative to deal with the measurement problem
in quantum mechanics.

But it turns out (see, e.g.,  \cite{hsy-jhep09,review4,hsy-jpcs12} and references therein) that
the complete formulation of the quantum aspects of spacetime requires a full-fledged quantum theory
from the start. In order to get a correct picture on the quantum origin of spacetime \cite{ncspace-ref,dfr-cmp95}, one cannot begin classically and then undergo quantization
in the traditional mold. (A similar viewpoint for the complete formulation of string/M-theory
was emphasized too in the Chapter 15 of Ref. \cite{b.greene}.)
It is a widely accepted consensus \cite{ncft1,ncft2} that, in a microscopic scale such
as the Planck scale $L_P \sim 10^{-33} {\rm cm}$ where the quantum effects of spacetime become important,
spacetime is no longer commuting but becomes noncommutative (NC), i.e.,
\begin{equation}\label{nc-space}
    [y^\mu, y^\nu] = i \theta^{\mu\nu}.
\end{equation}
In this paper we will consider only the Moyal-type noncommutativity where $\theta^{\mu\nu}$ is constant.
Although other NC spaces are of course possible, the Moyal NC space (\ref{nc-space}) will be enough
for our purpose since we will regard a general NC space as a (large) deformation of the Moyal NC space
due to NC gauge fields. According to the above philosophy, one has to regard the Heisenberg algebra (\ref{nc-space}) as a raw precursor to the fabric of spacetime which will be coalesced into an organized form
that we recognize as spacetime \cite{hsy-jhep09}. Unfortunately, the conventional wisdom is to interpret
the NC spacetime (\ref{nc-space}) as an extra structure (e.g., $B$-fields) defined on a preexisting spacetime. This description inevitably brings about the interpretation that the NC spacetime (\ref{nc-space})
necessarily breaks the Lorentz symmetry. This uneasy picture may be originated from the fact that
string theory is not a complete background independent formulation since our present formulation
of string theory presupposes the existence of space and time within which strings move about and vibrate.
(See the Chapter 15 of Ref. \cite{b.greene} for the vivid prospect of background independent formulation of string/M-theory.)

One of the reasons why one should not interpret the NC spacetime (\ref{nc-space}) as an extra structure defined on a preexisting spacetime is ironically coming from the string theory itself.
It is well-known \cite{string-book2} that open string theory admits a local gauge symmetry, the so-called
$\Lambda$-symmetry, defined by
\begin{equation} \label{lambda-symm}
(B,\; A) \to (B - d\Lambda, \; A + \Lambda)
\end{equation}
where the gauge parameter $\Lambda$ is a one-form in $M$. The $\Lambda$-symmetry is present only
when $B \neq 0$ and so it is a stringy symmetry in nature. When $B=0$, the symmetry (\ref{lambda-symm})
is reduced to $A \to A + d\lambda$, which is the ordinary $U(1)$ gauge symmetry.
The above local gauge symmetry in string theory must also be realized as the symmetry of
low energy effective theory. It turns out to be the case \cite{string-book2} that the low energy
effective field theory known as the Dirac-Born-Infeld (DBI) action really respects the local
symmetry (\ref{lambda-symm}). An essential point is that $\Lambda$-symmetry (\ref{lambda-symm}) can be considered as par with diffeomorphisms. This fact can be understood as follows \cite{sg-book1,sg-book2}. Suppose that the $B$-field in Eq. (\ref{lambda-symm}) is a symplectic structure on $M$, i.e.,
a nondegenerate, closed 2-form. For that case the symplectic structure $B$ defines a bundle
isomorphism $B : TM \to T^* M$ by $X \mapsto \Lambda = - \iota_X B$ where $\iota_X$ is an interior
product with respect to a vector field $X \in \Gamma(TM)$. Then the $\Lambda$-transformation
in Eq. (\ref{lambda-symm}) can be represented by $B' = B - d\Lambda = B + {\cal L}_X B$
where ${\cal L}_X$ is a Lie derivative along the flow of $X$. This means that the $\Lambda$-transformation
can be identified with a coordinate transformation generated by the vector field $X$.
(See Eq. (23) in Ref. \cite{review4} for an explicit verification.)
This fact elucidates why $\Lambda$-symmetry (\ref{lambda-symm}) can be regarded
as another independent diffeomorphism symmetry \cite{hsy-jhep09,cornalba,hsy-ijmp09}.
However this level of symmetry can be achieved only
when the $B$-field is present and so the $B$-field greatly enhances the underlying local gauge symmetry,
which is unprecedented in theories of particle physics such as the Standard Model.
Therefore it should be interesting to ponder on a physical consequence for the enhancement of
local gauge symmetry since a similar symmetry enhancement also arises in the presence of gravity.

It would be necessary to carefully contemplate our conventional wisdom imbued with any physical theory
all of which describe what happens in a given spacetime. In this mundane picture,
the NC spacetime (\ref{nc-space}) is interpreted as an extra structure induced by $B$-fields condensed
on a preexisting spacetime and so necessarily breaks the Lorentz symmetry.
But, as we emphasized before, the presence of $B$-fields rather introduces a local diffeomorphism
symmetry (\ref{lambda-symm}) which is not present in ordinary field theories (without $B$-fields).
Hence we have to ruminate on what had happened in the NC spacetime (\ref{nc-space}). Indeed the enhanced gauge symmetry when $B \neq 0$ gives us a hunch that there will be a radical change of physics--a new physics
in NC spacetime. Recently it was shown \cite{hsy-jhep09,review4,hsy-jpcs12} that the electromagnetism in NC spacetime can be realized as a theory of gravity and the symplectization of spacetime geometry is the origin of gravity. Remarkably the so-called emergent gravity reveals a novel picture about the origin of spacetime,
dubbed as emergent spacetime, which is radically different from the orthodox picture in general relativity.
See also related works in Refs. \cite{rivelles,review1,review2,review3,yasi-prd10,lee-yang,review6,review7}.

We believe that such a fallacy about NC spacetime hinders the view of the revolutionary aspects of emergent spacetime. In order to appreciate the notion of emergent gravity and correctly contrive quantum gravity
based on it, it would be worthwhile to explicitly show with some examples how the emergent gravity works.
In this paper we will examine a tiny yet circumstantial test of the emergent gravity picture with explicit
solutions in gravity and gauge theory. As a bottom-up approach of emergent gravity recently formulated
by us in Ref. \cite{our-jhep12}, we will derive symplectic $U(1)$ gauge fields starting from the Eguchi-Hanson metric \cite{eh-plb,eh-ap} in four-dimensional Euclidean gravity and show that they precisely reproduce $U(1)$
gauge fields of the Nekrasov-Schwarz instanton \cite{ns-inst} derived in Refs. \cite{sty-06,prl-06}.
As a top-down approach of emergent gravity, we take the $U(1)$ instanton found by Braden and
Nekrasov \cite{bn-inst} and derive a corresponding gravitational metric.
We will study the geometrical properties of the four-manifolds determined by the $U(1)$ instantons.

The paper is organized as follows. In Sec. II, we briefly explain how the emergent gravity picture arises
from the commutative description of NC gauge theory via the Seiberg-Witten (SW) map \cite{ncft-sw}.
This section is mostly devoted to fix the notations which will be used in later sections.
In Sec. III, we consider both the bottom-up and top-down approaches of emergent gravity as stated above.
We show that the gravitational metric of Braden-Nekrasov instanton exhibits a spacetime
singularity although it becomes a regular solution from the gauge theory point of view
after a K\"ahler blow up \cite{bn-inst}. In Sec. IV, we display an analysis of the relationship
between topological invariants associated with U(1) instantons and those of emergent four-dimensional
Riemannian manifolds. In Sec. V, we summarize the results obtained in this paper and
prove the formula (\ref{nc-curved}) for generic NC gauge fields.
In two appendices, we present the definition and several identities for 't Hooft symbols \cite{t-symbol1,t-symbol2,loy} and the explicit forms about the spin connections
and curvature tensors of a Riemannian metric we use in this paper.

\section{Emergent gravity}

It was shown in Refs. \cite{ncft-sw,dorn,jsw01,hsy-mpla06,ban-yan}, for slowly varying fields on a single
D-brane, that the dual description of the NC DBI action through the exact SW map is simply given by the ordinary DBI action expressed in terms of open string variables:
\begin{equation}\label{sw-equiv}
 \int d^4 y \sqrt{\det \big(G^{op}
  + \kappa (\Phi + \widehat{F}) \big)} = \int d^4 x \sqrt{\det{(1+ F \theta})}
 \sqrt{\det{ \big(G^{op} + \kappa (\Phi + {\bf F}) \big)}} +   {\cal O}(l_s \partial F),
\end{equation}
where $\kappa \equiv 2 \pi \alpha' = 2 \pi l_s^2$ and
\begin{equation}\label{def-fatf}
    {\bf F}_{\mu\nu} (x) \equiv \Bigl(\frac{1}{1 + F\theta} F \Bigr)_{\mu\nu} (x)
\end{equation}
with the ordinary $U(1)$ field strength defined by
\begin{equation}\label{cf}
F_{\mu\nu} (x) = \partial_\mu A_\nu(x) -  \partial_\nu A_\mu(x).
\end{equation}
Here slowly varying fields on a D-brane means symplectic gauge fields defined by the commutative description
of NC gauge fields and the field strength of symplectic gauge fields is given by
\begin{equation} \label{symp-f}
 \widehat{F}_{\mu\nu} (y) = \partial_\mu \widehat{A}_\nu (y)
- \partial_\nu \widehat{A}_\mu (y) + \{\widehat{A}_\mu, \widehat{A}_\nu \}_\theta (y).
\end{equation}
By comparing both sides of Eq. (\ref{sw-equiv}), one can immediately get the relation between commutative
and NC fields given by
\begin{eqnarray} \label{eswmap}
    &&  \widehat{F}_{\mu\nu}(y) = \Bigl(\frac{1}{1 + F\theta} F \Bigr)_{\mu\nu}(x), \\
    \label{measure-sw}
    && d^{4} y = d^{4} x \sqrt{\det(1+ F \theta)}(x),
\end{eqnarray}
where
\begin{equation}\label{cov-cod}
    x^\mu(y) \equiv y^\mu + \theta^{\mu\nu} \widehat{A}_\nu(y).
\end{equation}

An interesting point is that the SW equivalence (\ref{sw-equiv}) between commutative and
NC DBI actions\footnote{If the two descriptions are equivalent, the NC action defined by the left-hand
side of Eq. (\ref{sw-equiv}) must also respect two local gauge symmetries which
correspond to a NC version of the diffeomorphism symmetry and the $\Lambda$-symmetry (\ref{lambda-symm}).
The diffeomorphism symmetry may be more accessible by writing the determinant in the action (\ref{sw-equiv})
as $\det G^{op} \exp [ \sum_{n=1}^\infty \frac{(-1)^{n+1}(2 \pi \alpha')^n}{n}
\mathrm{Tr}\widehat{ \mathcal{F}}^n]$ where $\widehat{\mathcal{F}}_\mu^{~\nu} \equiv
(\Phi + \widehat{F})_{\mu \lambda} G_{op}^{\lambda\nu}$.
It is obvious that $\mathrm{Tr}\widehat{ \mathcal{F}}^n$ transforms as $\mathrm{Tr} \Xi \star
\widehat{ \mathcal{F}}^n \star \Xi^{-1}$ under diffeomorphism
$\Xi \in \mathrm{Diff}(M)$ and so it is invariant under the integral.
Thus the square root of the determinant in the action (\ref{sw-equiv}) has the transformation property
of a scalar density under NC diffeomorphisms \cite{nc-gravity}.
The $\Lambda$-symmetry (\ref{lambda-symm}) can be realized with a one-form $\widehat{\Lambda} = \widehat{\Lambda}_\mu (y) dy^\mu$ given by the transformation: $(\Phi, \widehat{A}) \to (\Phi - \widehat{D}\widehat{\Lambda} - i\widehat{\Lambda} \wedge \widehat{\Lambda}, \widehat{A} + \widehat{\Lambda})$
where $\widehat{D}\widehat{\Lambda} \equiv d\widehat{\Lambda} - i(\widehat{A} \wedge \widehat{\Lambda} + \widehat{\Lambda} \wedge \widehat{A})$ and the star-product is implicitly assumed for all formulas.
The NC $U(1)$ gauge transformation then corresponds to a special case of the NC $\Lambda$-symmetry with $\widehat{\Lambda}_\mu = \widehat{D}_\mu \widehat{\lambda}$
while ignoring nonlinear terms $[\widehat{\Lambda}_\mu, \widehat{\Lambda}_\nu]_\star$.}
can be derived using only an elementary property, known as the Darboux theorem or the Moser lemma \cite{sg-book1,sg-book2}, in symplectic geometry. Indeed the derivation is based on several important
pictures for emergent gravity. First of all, the $\Lambda$-symmetry (\ref{lambda-symm}) enforces
the gauge invariant quantity as the form $\mathcal{F} = B + F$ where $F = dA$.
Consequently the dynamical gauge fields $A_\mu (x)$ fluctuating on a symplectic manifold $(M,B)$ manifest themselves only as a deformation of the underlying symplectic structure $B$. The Darboux theorem or
the Moser lemma in symplectic geometry then implies that it is always possible to find a local coordinate transformation to eliminate the electromagnetic force $F=dA$ in the total field
strength ${\mathcal F} = B + F$. In other words, as long as the space $M$ admits a symplectic structure,
one can find a local coordinate transformation $\phi: x \mapsto y=y(x)$ on $U \subset M$ \cite{cornalba}
such that
\begin{equation}\label{darboux-tr}
 \Big(B_{ab} + F_{ab}(x) \Big)\frac{\partial x^a}{\partial y^\mu}
 \frac{\partial x^b}{\partial y^\nu} = B_{\mu\nu}.
\end{equation}
By taking the inverse of Eq. (\ref{darboux-tr}), one can rewrite Eq. (\ref{darboux-tr}) in the form
\begin{equation}\label{poisson-tr}
    \Theta^{ab} (x) \equiv \Big( \frac{1}{B+F} \Big)^{ab}(x) =
  \theta^{\mu\nu}  \frac{\partial x^a}{\partial y^\mu}
 \frac{\partial x^b}{\partial y^\nu} = \{ x^a, x^b \}_\theta (y).
\end{equation}
Using the representation (\ref{cov-cod}) for the coordinate transformation $x^a = x^a (y)$,
Eq. (\ref{poisson-tr}) reads as
\begin{equation}\label{sw-map}
\Theta^{ab} (x) = \Big(\theta - \theta  \widehat{F} \theta
\Big)^{ab} (y) \quad \Leftrightarrow \quad
\widehat{F}_{\mu\nu} (y) = \Big(\frac{1}{1+ F \theta}
F\Big)_{\mu\nu} (x).
\end{equation}
Then Eq. (\ref{measure-sw}) is simply the Jacobian $J=|\frac{\partial y}{\partial x}| = \sqrt{\det(1+ F \theta)}$ of the coordinate transformation $x \mapsto y=y(x)$ which can be derived from Eq. (\ref{darboux-tr})
by taking the determinant on both sides.

Consequently one can see that the SW map in Eqs. (\ref{eswmap}) and (\ref{measure-sw})
can be obtained by the coordinate transformation (\ref{darboux-tr}) that locally eliminates
the electromagnetic force $F=dA$ \cite{hsy-jhep09,hsy-ijmp09,ban-yan}. In fact,
the coordinate transformation (\ref{darboux-tr}) can be understood as the $\Lambda$-transformation
or $B$-field transformation, $B \to B' = B - d\Lambda$, with $\Lambda = - A$ in Eq. (\ref{lambda-symm}).
As we emphasized in Sec. I, the $B$-field transformation
can be realized as a diffeomorphism $\phi: M \to M$ generated by a vector field $X$ obeying $A = \iota_X B$
and so $F = dA = \mathcal{L}_X B$. Hence the coordinate transformation (\ref{darboux-tr}) forms
a one-parameter group of diffeomorphisms generated by the flow along $X$ \cite{sg-book1,sg-book2}.
In the end there exists a novel form of the equivalence principle \cite{hsy-jhep09,hsy-ijmp09} such that
the electromagnetic force can always be eliminated by a local coordinate transformation as long as
$U(1)$ gauge theory is defined on a symplectic manifold $(M, B)$.
A striking picture then comes out \cite{hsy-jhep09,review4,hsy-jpcs12} that gravity can emerge
from NC $U(1)$ gauge theory as a natural result of the equivalence principle for the electromagnetic force.

Now we will illuminate how the emergent gravity picture stems from the SW equivalence (\ref{sw-equiv})
which will also serve to set up the notations used later.
We assume the open string metric $G^{op}_{\mu\nu} = \delta_{\mu\nu}$ for simplicity.
One can expand both sides of Eq. (\ref{sw-equiv}) into power series of $\kappa$.
At $\mathcal{O}(\kappa^2)$ one can get the following identity
\begin{equation}\label{second-sw-equiv}
    \frac{1}{4} \int d^4 y \mathrm{Tr}(\widehat{F} + \Phi)^2
    = \frac{1}{4} \int d^4 x \sqrt{G} \mathrm{Tr} (\mathbf{F} + \Phi)^2,
\end{equation}
where $\mathrm{Tr}(AB) = A_{\mu\nu} B_{\nu\mu}$ and we introduced an effective metric
\begin{equation}\label{eff-metric}
    G_{\mu\nu} \equiv \delta_{\mu\nu} + (F\theta)_{\mu\nu}, \qquad
     G^{\mu\nu} \equiv (G^{-1})^{\mu\nu} = \Big( \frac{1}{1 + F\theta} \Big)^{\mu\nu}
\end{equation}
determined by $U(1)$ gauge fields. The ``effective metric" (\ref{eff-metric}) emergent
from $U(1)$ gauge fields is in general not symmetric because $G - G^T = F \theta - \theta F \neq 0$.
In four dimensions, the six-dimensional vector space $\Lambda^2 T^* M$ of
two-forms splits canonically into the sum of three-dimensional vector spaces of self-dual and anti-self-dual
two-forms and also the six-dimensional vector space $\Lambda^2 TM$ of bi-vectors splits similarly.
So let us take the following decompositions:
\begin{eqnarray} \label{hodge-f}
F_{\mu\nu} &=& f^{(+)i} \eta^i_{\mu\nu} + f^{(-)i} \overline{\eta}^i_{\mu\nu}, \\
\label{hodge-theta}
\theta^{\mu\nu} &=& \theta^{(+)i} \eta^i_{\mu\nu} + \theta^{(-)i} \overline{\eta}^i_{\mu\nu},
\end{eqnarray}
where $\eta^i_{\mu\nu}$ and $\overline{\eta}^i_{\mu\nu}\; (i=1,2,3)$ are self-dual and anti-self-dual
't Hooft symbols, respectively. See the Appendix A for the definition and the properties of the 't Hooft symbols. A general condition for the metric (\ref{eff-metric}) to be symmetric is given by
\begin{equation}\label{g-symm}
    \varepsilon^{ijk}  f^{(+)j} \theta^{(+)k} = 0 = \varepsilon^{ijk}  f^{(-)j} \theta^{(-)k}, \quad
    \forall i =1,2,3.
\end{equation}
This means that $F_{\mu\nu}$ and $\theta^{\mu\nu}$ being the second rank tensors of
$SO(4) = SU(2)_L \times SU(2)_R$ are parallel to each other in the vector space of $su(2)_L$ and
$su(2)_R$ Lie algebras. In this case the metric (\ref{eff-metric}) becomes symmetric, i.e.
$G = G^T$ and so it can be regarded as a usual Riemannian metric. We will implicitly assume
the condition (\ref{g-symm}), otherwise we will simply consider the effective metric (\ref{eff-metric})
as a notation for a specific form of $U(1)$ gauge fields.

In the usual NC description with $\Phi = 0$, the identity (\ref{second-sw-equiv}) takes the form \cite{hsy-mpla06,ban-yan}
\begin{equation}\label{sw-equiv-0}
    \frac{1}{4} \int d^4 y \widehat{F}_{\mu\nu}  \widehat{F}^{\mu\nu}
    = \frac{1}{4} \int d^4 x \sqrt{G} G^{\mu\rho}G^{\sigma\nu} F_{\mu\nu} F_{\rho\sigma},
\end{equation}
while, in the background independent prescription with $\Phi = -B$ \cite{ncft-sw,nc-seiberg},
the identity (\ref{second-sw-equiv}) can be written in the form
\begin{equation}\label{sw-equiv-b}
    \frac{1}{4} \int d^4 y \{C_\mu, C_\nu \}_\theta^2
    = \frac{1}{4} \int d^4 x \sqrt{G} G^{\mu\rho}G^{\sigma\nu} B_{\mu\nu} B_{\rho\sigma},
\end{equation}
where $C_\mu (y) \equiv B_{\mu\nu} x^\nu (y) = B_{\mu\nu} y^\nu + \widehat{A}_\mu (y)$ and
we used the relation
\begin{equation}\label{cc-poisson}
    \{C_\mu, C_\nu \}_\theta = - B_{\mu\nu} + \widehat{F}_{\mu\nu}.
\end{equation}
One can see that the dual description of NC $U(1)$ gauge theory via the SW map can be interpreted
as the ordinary Maxwell theory coupling to the effective metric (\ref{eff-metric}) determined
by $U(1)$ gauge fields \cite{rivelles}. In particular, the background independent description (\ref{sw-equiv-b}) clearly shows that the fluctuations of NC photons around the background $B$-field
are mapped through the SW map to the fluctuations of spacetime geometry.

The field strength of NC $U(1)$ gauge fields is defined by quantizing the semi-classical version
(\ref{symp-f}) and it is given by
\begin{equation} \label{nc-f}
 \widehat{F}_{\mu\nu}  = \partial_\mu \widehat{A}_\nu - \partial_\nu \widehat{A}_\mu
 -i  [\widehat{A}_\mu, \widehat{A}_\nu ]_\star.
\end{equation}
Since the NC field strength is nonlinear due to the commutator term, one can consider a nontrivial
solution of the following self-duality equation \cite{ns-inst,nek-cmp03,furu-ptp,furu-ptps,kly-jkps02,kly-plb01}
\begin{equation}\label{self-dual-inst}
 \widehat{F}_{\mu\nu} (y) = \pm \frac{1}{2} {\varepsilon_{\mu\nu}}^{\rho\sigma}
 \widehat{F}_{\rho\sigma} (y).
\end{equation}
A solution of the self-duality equation (\ref{self-dual-inst}) is called a NC $U(1)$ instanton whereas
it will be called a symplectic $U(1)$ instanton for the semi-classical limit where the $U(1)$ field
strength is defined by Eq. (\ref{symp-f}). But we can apply the commutative description
to NC $U(1)$ instantons using the identity (\ref{sw-equiv-0}).
Using the property $\mathbf{F}_{\mu\nu} = - \mathbf{F}_{\nu\mu}$, it is easy to rewrite the right-hand
side of Eq. (\ref{sw-equiv-0}) in the Bogomolnyi form \cite{bps-paper}
\begin{equation}\label{bogomolnyi1}
    S_C = \frac{1}{8} \int d^4 x \sqrt{G} \Bigl( {\bf F}_{\mu\nu} \mp \frac{1}{2}
{\varepsilon_{\mu\nu}}^{\rho\sigma} {\bf F}_{\rho\sigma} \Bigr)^2
\pm \frac{1}{8} \int d^4 x \varepsilon^{\mu\nu\rho\sigma} F_{\mu\nu} F_{\rho\sigma}.
\end{equation}
The Bogomolnyi form (\ref{bogomolnyi1}) immediately shows that the first term is positive definite
while the second term is topological, i.e. a boundary term and thus does not affect
the equations of motion. Hence the minimum of the action $S_{\mathrm{C}}$ is achieved
in the configurations satisfying the self-duality equation \cite{sty-06}
\begin{equation} \label{c-self-dual}
{\bf F}_{\mu\nu} (x) = \pm \frac{1}{2} {\varepsilon_{\mu\nu}}^{\rho\sigma} {\bf F}_{\rho\sigma} (x).
\end{equation}
Note that the above equation is directly obtained by applying the exact SW map (\ref{eswmap}) (in
the semi-classical limit) to the NC self-duality equation (\ref{self-dual-inst}).
A solution obeying the self-duality equation (\ref{c-self-dual}) was dubbed above a
symplectic $U(1)$ instanton as a commutative limit of NC $U(1)$ instanton.

We will take the NC space (\ref{nc-space}) as self-dual that means $\theta^{(-)i} = 0$
in Eq. (\ref{hodge-theta}). It is always possible to rotate $\theta^{(+)i}$ into $(0,0, \theta^{(+)3})$
such that $\theta^{\mu\nu} = \frac{\theta}{2} \eta^3_{\mu\nu}$.
In that case the condition (\ref{g-symm}) can be satisfied
if $f^{(+)1} = f^{(+)2} = 0$ and the $U(1)$ field strength (\ref{hodge-f}) then takes the form
\begin{equation} \label{symg-f}
F_{\mu\nu} = f^{(+)3} \eta^3_{\mu\nu} + f^{(-)i} \overline{\eta}^i_{\mu\nu}.
\end{equation}
Such $U(1)$ gauge fields result in a usual Riemannian metric \cite{sty-06,prl-06}.
Therefore one can view the right-hand side of Eq. (\ref{sw-equiv-0}) as $U(1)$ gauge theory defined on
a Riemannian manifold whose metric is given by Eq. (\ref{eff-metric}).\footnote{However,
it should not be interpreted as a gauge theory defined on a fixed background manifold because
the four-dimensional metric (\ref{eff-metric}) depends in turn on dynamical $U(1)$ gauge fields.}
Hence one can derive the Bogomolnyi bound for the right-hand side of Eq. (\ref{sw-equiv-0}) exactly
in the same way as a gauge theory defined on a curved manifold:
\begin{widetext}
\begin{eqnarray}\label{bogomolnyi2}
    S_C &=& \frac{1}{8} \int d^4 x \sqrt{G} G^{\mu\rho} G^{\nu\sigma} \Bigl(F_{\mu\nu} \mp \frac{1}{2}
\frac{\varepsilon^{\lambda\tau\alpha\beta}}{\sqrt{G}} G_{\mu\lambda} G_{\nu\tau} F_{\alpha\beta} \Bigr)
\Bigl(F_{\rho\sigma} \mp \frac{1}{2} \frac{\varepsilon^{\xi\eta\gamma\delta}}{\sqrt{G}}
G_{\rho\xi} G_{\sigma\eta} F_{\gamma\delta} \Bigr) \nonumber \\
&& \pm \frac{1}{8} \int d^4 x \varepsilon^{\mu\nu\rho\sigma} F_{\mu\nu} F_{\rho\sigma}.
\end{eqnarray}
\end{widetext}
Accordingly, the self-duality equation for the action $S_C$ is now given by \cite{hsy-ijmp09}
\begin{equation} \label{curved-sde}
F_{\mu\nu} = \pm \frac{1}{2} \frac{\varepsilon^{\lambda\tau\rho\sigma}}{\sqrt{G}}
G_{\mu\lambda} G_{\nu\tau} F_{\rho\sigma}.
\end{equation}
This equation suggests that symplectic $U(1)$ instantons can be interpreted as (anti-)self-dual
$U(1)$ connections on a four-manifold whose metric is given by (\ref{eff-metric}).
Actually if we introduce vierbeins of the metric (\ref{eff-metric}) such that
\begin{equation}\label{eff-vierbein}
    ds^2 = G_{\mu\nu}(x) dx^\mu \otimes dx^\nu = E^a \otimes E^a,
\end{equation}
the above self-duality equation (\ref{curved-sde}) can be written as
\begin{equation}\label{form-sde}
    F = \pm *F
\end{equation}
where $F = \frac{1}{2} F_{ab} E^a \wedge E^b$ and $*$ denotes the Hodge dual operation on forms.
Or, in the component form, Eq. (\ref{form-sde}) reads as
\begin{equation}\label{sde-comp}
    F_{ab} = \pm \frac{1}{2} {\varepsilon_{ab}}^{cd} F_{cd}.
\end{equation}
It is easy to show that Eq. (\ref{sde-comp}) is equivalent to Eq. (\ref{curved-sde})
using the definition $F_{ab} = E^\mu_a E^\nu_b F_{\mu\nu}$.

A similar argument can be applied to the background independent description (\ref{sw-equiv-b}) although
the action (\ref{sw-equiv-b}) diverges in general. One can introduce a regularized action by subtracting
the most divergent piece and define the theory with the action
\begin{equation}\label{compact-action}
S_R = \frac{1}{4} \int d^4 x \sqrt{G} G^{\mu\rho}G^{\sigma\nu} B_{\mu\nu} B_{\rho\sigma}
- \frac{1}{4} \int d^4 x B^2_{\mu\nu}.
\end{equation}
Note that the subtraction does not affect the equations of motion and the above regularized action
becomes finite. One can then implement the Bogomolyni bound to the regularized action (\ref{compact-action})
and the result is simply given by
\begin{eqnarray}\label{bogomolnyi3}
    S_R &=& \frac{1}{8} \int d^4 x \sqrt{G} G^{\mu\rho} G^{\nu\sigma} \Bigl(B_{\mu\nu} \mp \frac{1}{2}
\frac{\varepsilon^{\lambda\tau\alpha\beta}}{\sqrt{G}} G_{\mu\lambda} G_{\nu\tau} B_{\alpha\beta} \Bigr)
\Bigl(B_{\rho\sigma} \mp \frac{1}{2} \frac{\varepsilon^{\xi\eta\gamma\delta}}{\sqrt{G}}
G_{\rho\xi} G_{\sigma\eta} B_{\gamma\delta} \Bigr) \nonumber \\
&& - \frac{1}{4} \int d^4 x \Bigl(B_{\mu\nu} \mp \frac{1}{2} {\varepsilon_{\mu\nu}}^{\rho\sigma}
B_{\rho\sigma} \Bigr) B^{\mu\nu}.
\end{eqnarray}
This procedure thus leads to another form of the self-duality equation
\begin{equation} \label{curved-sde2}
B_{\mu\nu} = \pm \frac{1}{2} \frac{\varepsilon^{\lambda\tau\rho\sigma}}{\sqrt{G}}
G_{\mu\lambda} G_{\nu\tau} B_{\rho\sigma}
\end{equation}
which is equivalent, in terms of form language, to
\begin{equation}\label{b-sde}
    B = \pm *B
\end{equation}
with $B = \frac{1}{2} B_{ab} E^a \wedge E^b$. Note that the second term in Eq. (\ref{bogomolnyi3})
is a total derivative term because $B = dA^{(0)}$ with $A_\mu^{(0)} = - \frac{1}{2}B_{\mu\nu} x^\nu$
and so a boundary term on $\mathbb{S}^3 = \partial \mathbb{R}^4$ at $|x| \to \infty$.
At the asymptotic region, $|x| \to \infty$, where the metric $G_{\mu\nu}$ reduces to $\delta_{\mu\nu}$,
the self-duality equation (\ref{curved-sde2}) reduces to $B_{\mu\nu} = \pm \frac{1}{2} {\varepsilon_{\mu\nu}}^{\rho\sigma} B_{\rho\sigma}$ and so the last term in $S_R$ identically vanishes.

As was shown before, the SW equivalence (\ref{sw-equiv}) is based on a novel form of
the equivalence principle for the electromagnetic force. The quantization of the symplectic
manifold $(M, B)$ brings about the NC spacetime (\ref{nc-space}) and results in NC $U(1)$ gauge theory. Consequently, the equivalence principle for the electromagnetic force guarantees that (quantum) gravity
can emerge from NC $U(1)$ gauge theory \cite{hsy-jhep09}. If so,
a natural question is what kind of four-manifold arises from a solution of the self-duality equation
(\ref{self-dual-inst}) known as NC $U(1)$ instantons \cite{ns-inst}. In this paper we will focus on its commutative limit satisfying the self-duality equation (\ref{c-self-dual}) called symplectic
$U(1)$ instantons. We showed that the self-duality equation of symplectic $U(1)$ instantons
can be written in the form (\ref{curved-sde}) since the metric (\ref{eff-metric})
for the solution of Eq. (\ref{c-self-dual}) is symmetric \cite{sty-06,prl-06}.
It was shown \cite{sty-06,prl-06,hsy-epl09} that the equation (\ref{c-self-dual}) describes gravitational
instantons obeying the self-dual equations \cite{egh-report,besse}
\begin{equation}\label{g-inst}
    R_{abef} = \pm \frac{1}{2}{\varepsilon_{ab}}^{cd} R_{cdef},
\end{equation}
where $R_{abcd}$ is a Riemann curvature tensor. More precisely, if one identifies from
the effective metric (\ref{eff-metric}) a gravitational metric defined by
\begin{equation}\label{G-g}
    G_{\mu\nu} (x) = \frac{1}{2} ( \delta_{\mu\nu} + g_{\mu\nu}(x) ),
\end{equation}
the metric $g_{\mu\nu}(x)$ describes a Ricci-flat K\"ahler manifold obeying Eq. (\ref{g-inst}).
In other words, the four-manifold whose metric is given by
\begin{equation}\label{riem-metric}
    ds^2 = g_{\mu\nu} (x) dx^\mu \otimes dx^\nu = e^a \otimes e^a
\end{equation}
is a hyper-K\"ahler manifold \cite{hsy-epl09}. In next section we will verify the emergent gravity picture
with explicit solutions from both the bottom-up and the top-down approaches.

\section{Four-manifolds and $U(1)$ gauge fields}

In Sec. II, we explained why the Riemannian metric (\ref{riem-metric}) can arise from $U(1)$ gauge fields
on a symplectic manifold $(M,B)$ and how it can be determined by solving the equations
of motion for the $U(1)$ gauge fields. But the emergent gravity picture can be inverted,
as recently formulated in Ref. \cite{our-jhep12}, such that one gets $U(1)$ gauge fields using the relation (\ref{eff-metric}) whenever a Riemannian metric $(M, g)$ is given.
Now we will illustrate how the emergent gravity works for both the top-down and the bottom-up approaches.
For that purpose, we will take an explicit solution in general relativity whose metric is assumed
to be of the form
\begin{equation}\label{metric-form}
    ds^2 = A^2(r) (dr^2 + r^2 \sigma_3^2) + B^2(r) r^2(\sigma_1^2 + \sigma_2^2)
\end{equation}
and so the covectors (vierbeins) are given by
\begin{equation}\label{co-vector}
    e^1 = B(r) r \sigma^1, \quad e^2 = B(r) r \sigma^2, \quad e^3 = A(r) r \sigma^3, \quad
    e^4 = A(r) dr.
\end{equation}
We have introduced a left-invariant coframe $\{ \sigma^i: i =1,2,3 \}$ for $\mathbb{S}^3$ defined by
\cite{egh-report}
\begin{equation}\label{coframe-s3}
    \sigma^i = - \frac{1}{r^2} \eta^i_{\mu\nu} x^\mu dx^\nu
\end{equation}
where $r^2 = x_1^2 + \cdots + x_4^2$. They obey the following structure equations
\begin{equation}\label{st-eqs3}
    d\sigma^i = -\varepsilon^{ijk} \sigma^j \wedge \sigma^k.
\end{equation}
In Appendix B, we present the explicit results for the spin connections and curvature tensors
determined by the metric (\ref{metric-form}).

We will assume that the metric (\ref{metric-form}) is asymptotically locally Euclidean (ALE),
i.e., $A(r) = B(r) \to 1$ as $r \to \infty$. In that case, it will be useful to introduce the Hopf map $\pi: \mathbb{S}^3 \to \mathbb{S}^2$ which can be represented in terms of $\mathbb{R}^4$ variables as \cite{sty-06}
\begin{eqnarray}\label{hopf}
      && T^1 = - (x^1 x^3 + x^2 x^4), \nonumber \\
      && T^2 = x^1 x^4 - x^2 x^3, \\
      && T^3 = \frac{1}{2} (x_1^2 + x^2_2 - x_3^2 - x_4^2) \nonumber
\end{eqnarray}
and
\begin{equation}\label{t-square}
    \sum_{i=1}^3 T^i T^i = \frac{r^4}{4}.
\end{equation}
The following relations may be useful for later purpose (see Eq. (3.32) in Ref. \cite{sty-06}):
\begin{equation}\label{id-t}
    \overline{\eta}^i_{\mu\nu} \partial_\nu T^i = 3 \eta^3_{\mu\nu} x^\nu,
    \qquad \overline{\eta}^i_{\mu\nu} x^\nu T^i = \frac{r^2}{2} \eta^3_{\mu\nu} x^\nu.
\end{equation}

\subsection{$U(1)$ instanton from Eguchi-Hanson metric}

The Eguchi-Hanson metric \cite{eh-plb,eh-ap} describes a non-compact, self-dual,
ALE space on the cotangent bundle of 2-sphere $T^* \mathbb{S}^2$
with SU(2) holonomy group. The explicit form of the metric is given by
\begin{equation}\label{eh-metric}
    ds^2 = f^{-1}(\rho) d\rho^2 + \rho^2 \bigl(\sigma_1^2 + \sigma_2^2 + f(\rho) \sigma_3^2 \bigr)
\end{equation}
where $f(\rho) = 1- \frac{t^4}{\rho^4}$ and $\rho^4 = r^4 + t^4$.
Thus the Eguchi-Hanson metric takes the form (\ref{metric-form}) with
\begin{equation}\label{metric-ab}
    A^2(r) = \frac{r^2}{\sqrt{r^4 + t^4}} = B^{-2}(r).
\end{equation}
In order to write the metric in terms of the Cartesian coordinates $\{ x^\mu \}$,\footnote{\label{caveat}In order to
avoid a confusion, we want to point out that the Cartesian coordinates $\{ x^\mu \}$ in the one-form (\ref{coframe-s3}) and the Hopf map (\ref{hopf}) should be regarded as the coordinates on the flat space $\mathbb{R}^4$. Therefore it is not necessary to concern about raising and lowering the indices
$\mu, \nu, \cdots$ in the one-forms $(\sigma^i, dr = \frac{x^\mu dx^\mu}{r})$ and the Hopf coordinates $T^i$.}
let us plug Eq. (\ref{coframe-s3}) in Eq. (\ref{eh-metric}). The result can be written as
\begin{eqnarray}\label{cart-eh}
    ds^2 &=& g_{\mu\nu} (x)  dx^\mu \otimes dx^\nu \nonumber \\
    &=& \Bigl[\frac{\sqrt{r^4 + t^4}}{2r^2} \bigl(f(r) + 1 \bigr) \delta_{\mu\nu}
    - \frac{\sqrt{r^4 + t^4}}{r^4} \bigl(f(r) - 1 \bigr)
    \bigl(\eta^3 \overline{\eta}^i \bigr)_{\mu\nu} T^i \Bigr] dx^\mu \otimes dx^\nu
\end{eqnarray}
after using the identity
\begin{equation}\label{id-hopf}
    x^\mu x^\nu + \eta^3_{\mu\rho} \eta^3_{\nu\sigma} x^\rho x^\sigma
    = \frac{r^2}{2} \delta^{\mu\nu} - \bigl(\eta^3 \overline{\eta}^i \bigr)_{\mu\nu} T^i
\end{equation}
which can be checked by a straightforward calculation. Later we will also use the following identity
\begin{equation}\label{id-bn}
    \eta^3_{\mu\rho} x^\rho x^\nu - \eta^3_{\nu\rho}x^\mu x^\rho
    = \frac{r^2}{2} \eta^3_{\mu\nu} + \overline{\eta}^i_{\mu\nu} T^i
\end{equation}
which can be derived from Eq. (\ref{id-hopf}) by multiplying $\eta^3_{\mu\rho}$.

Now it is straightforward to identify $U(1)$ gauge fields from the Eguchi-Hanson metric (\ref{cart-eh}).
Combining Eqs. (\ref{eff-metric}) and (\ref{G-g}) leads to the relation
\begin{equation}\label{g-f}
    g_{\mu\nu}(x) = \delta_{\mu\nu} + 2 (F\theta)_{\mu\nu}.
\end{equation}
For our choice $\theta^{\mu\nu} = \frac{\theta}{2} \eta^3_{\mu\nu}$ where we put $\theta=1$ for simplicity,
the metric (\ref{cart-eh}) leads to the $U(1)$ field strength
\begin{eqnarray}\label{u1-eh}
    F_{\mu\nu}(x) &=& \frac{t^4}{\rho^2 r^4} \overline{\eta}^i_{\mu\nu} T^i
    - \frac{(\rho^2 - r^2)^2}{2 \rho^2 r^2} \eta^3_{\mu\nu} \nonumber \\
    &=& \frac{t^4}{r^6\sqrt{1 + \frac{t^4}{r^4}}} \overline{\eta}^i_{\mu\nu} T^i
    -\frac{\Bigl(\sqrt{1 + \frac{t^4}{r^4}} -1 \Bigr)^2}{2 \sqrt{1 + \frac{t^4}{r^4}}} \eta^3_{\mu\nu}.
\end{eqnarray}
The above result is exactly the same as the field strength of symplectic $U(1)$ gauge fields
(see Eq. (3.25) in Ref. \cite{sty-06}) determined by solving the self-duality equation (\ref{c-self-dual}).
It was shown in Refs. \cite{ncft-sw,sty-06} that the result (\ref{u1-eh}) can be obtained from the
commutative description of the Nekrasov-Schwarz instanton.
Therefore, starting from the Eguchi-Hanson metric (\ref{eh-metric}) in four-dimensional Euclidean gravity,
we precisely derived $U(1)$ gauge fields of the Nekrasov-Schwarz instanton and thus checked the bottom-up
approach of emergent gravity \cite{our-jhep12}. This fact can be further confirmed by calculating
the $U(1)$ field strength (\ref{symp-f}) using the exact SW-map (\ref{eswmap}):
\begin{equation}\label{esw-nsu1}
    \widehat{F}_{\mu\nu} (x) = \frac{4}{r^2} \frac{\sqrt{1 + \frac{t^4}{r^4}} -1}{\sqrt{1
    + \frac{t^4}{r^4}} + 1} \overline{\eta}^i_{\mu\nu} T^i
\end{equation}
which clearly satisfies the self-duality equation (\ref{self-dual-inst}) with $-$-sign.
In the end the bottom-up approach nicely verifies the result in Ref. \cite{sty-06} that
the Eguchi-Hanson metric (\ref{eh-metric}) comes from the NC $U(1)$ instanton satisfying
the self-duality equation (\ref{self-dual-inst}).

In Sec. II, we observed that the self-duality equation (\ref{self-dual-inst}) for NC $U(1)$ instantons
can be written in several equivalent forms, Eqs. (\ref{c-self-dual}), (\ref{curved-sde}) and
(\ref{curved-sde2}), in the commutative description after the SW map.
Considering the fact that they look quite different at first sight,
the existence of such equivalent statements is an interesting property.
In order to check the identities, first note the relation (\ref{G-g}) where the metric
$g_{\mu\nu}$ refers to Eq. (\ref{cart-eh}) and so the metric (\ref{eff-metric}) is represented by
\begin{equation}\label{metric-matrix}
    G_{\mu\nu} = \left(
                   \begin{array}{cccc}
                     G_1 & 0 & G_3 & G_4 \\
                     0 & G_1 & -G_4 & G_3 \\
                     G_3 & -G_4 & G_2 & 0 \\
                     G_4 & G_3 & 0 & G_2 \\
                   \end{array}
                 \right)
\end{equation}
where
\begin{eqnarray*}
   && G_1 = P - Q T^3, \qquad
    G_2 = P + Q T^3, \\
 && G_3 = Q T^1, \qquad \qquad G_4 = - Q T^2
\end{eqnarray*}
and
\begin{equation*}
    P = \frac{(r^2 + \sqrt{r^4 + t^4})^2}{4r^2\sqrt{r^4 + t^4}}, \qquad Q = \frac{t^4}{2 r^4 \sqrt{r^4 + t^4}}.
\end{equation*}
In a compact notation, the metric (\ref{metric-matrix}) can be written as
\begin{equation}\label{metric-cform}
    G_{\mu\nu}(x) = P \delta_{\mu\nu} + Q (\eta^3\overline{\eta}^i)_{\mu\nu} T^i.
\end{equation}
The next thing is to calculate the square root of $\det G_{\mu\nu}$ which reads as
\begin{equation}\label{det-g}
    \sqrt{G} = G_1 G_2 - (G_3^2 + G_4^2) = P^2 - \frac{r^4}{4} Q^2.
\end{equation}
It is now straightforward to check the self-duality equation (\ref{curved-sde}) (with $-$-sign)
using the results in Eqs. (\ref{u1-eh}) and (\ref{metric-matrix}).
It is amusing to see that the symplectic $U(1)$ gauge fields derived from the Eguchi-Hanson metric
manifestly become anti-self-dual with respect to the metric (\ref{eff-metric}) generated by themselves
while they are neither self-dual nor anti-self-dual with respect to the flat metric on $\mathbb{R}^4$
as one can see from Eq. (\ref{u1-eh}).

Note that Eq. (\ref{curved-sde}) takes exactly the same form as the self-duality equation
defined on a Riemannian manifold with the metric (\ref{eff-metric}).
Indeed we showed that Eq. (\ref{curved-sde}) can be cast into the form (\ref{sde-comp})
when we define $F_{ab} = E^\mu_a E^\nu_b F_{\mu\nu}$. In order to properly understand Eq. (\ref{sde-comp}),
we have to point out a caveat. So far it was not necessary to distinguish between the world (curved space)
indices $\mu, \nu, \cdots$ and frame (tangent space) indices $a,b,\cdots$. (See the footnote \ref{caveat}.)
Now, if one intends to interpret Eq. (\ref{curved-sde}) as the form (\ref{sde-comp}),
the $\mu, \nu$ indices in $F_{\mu\nu} = E_\mu^a E_\nu^b F_{ab}$ have to be regarded as
the world indices and so they must be raised and lowered using the metric $G_{\mu\nu}$
as in general relativity. If we adopt this interpretation, we get a remarkable
picture about NC gauge fields. A naive observation is the following. The self-duality equation (\ref{curved-sde}) says that the commutative field strength $F_{\mu\nu}$ is (anti-)self-dual
with respect to the metric $G_{\mu\nu}$. If we introduce a local basis $\{E_a\}$ for the tangent
bundle $TM$ and the dual basis $\{E^a \in T^*M \}$ defined by Eq. (\ref{eff-vierbein}),
the self-duality equation (\ref{curved-sde}) can be written in the form (\ref{sde-comp}) in a locally
inertial frame where $F_{ab}$ becomes (anti-)self-dual with respect to the flat metric $\delta_{ab}$.
We know that $\widehat{F}_{\mu\nu}$ in Eq. (\ref{esw-nsu1}) is anti-self-dual with respect
to the flat metric $\delta_{\mu\nu}$ and so it is natural to identify $F_{ab}$ with Eq. (\ref{esw-nsu1}).
This reasoning implies an intriguing relation
\begin{equation}\label{nc-curved}
 F_{ab} = E^{~\mu}_a F_{\mu\nu}  E^\nu_{~b} = \widehat{F}_{ab}
\end{equation}
where $\widehat{F}_{ab}$ is given by Eq. (\ref{esw-nsu1}) with the replacement $(\mu,\nu) \to (a,b)$.
Now we will prove the above identity.

It is easy to find the vierbeins $E_\mu^{~a}$ and the inverse vierbeins $E_a^{~\mu}$ from the metric (\ref{metric-cform}):
\begin{eqnarray} \label{vierbein}
&& E_\mu^{~a} = C \delta_\mu^{~a} + D (\eta^3 \overline{\eta}^i)_\mu^{~a} T^i, \\
\label{vierbein-1}
&& E_a^{~\mu} = \pm G^{-1/4} \bigl(C \delta_a^{~\mu} - D (\eta^3 \overline{\eta}^i)_a^{~\mu} T^i \bigr),
\end{eqnarray}
where $\sqrt{G}$ is given by Eq. (\ref{det-g}) and
\begin{equation}\label{cd-coeff}
    C^2 = \frac{1}{2} \bigl(P \pm G^{1/4} \bigr), \qquad
    D^2 = \frac{2}{r^4} \bigl(P \mp G^{1/4} \bigr).
\end{equation}
Here we understand the above matrix products as  $(AB)_\mu^{~a} = A_{\mu \lambda} B^{\lambda a}$
and $(AB)_a^{~\mu} = A_{a \lambda} B^{\lambda\mu}$.
We define Eqs. (\ref{eta^2}) and (\ref{eta-ex}) with the matrix product and used them to derive the above results.
Since $G_{\mu\nu} = \delta_{\mu\nu} + (F\theta)_{\mu\nu}$ in Eq. (\ref{metric-cform}),
one can represent the $U(1)$ field strength as
\begin{equation}\label{f-proof}
F_{\mu\nu} = 2 \bigl( Q\overline{\eta}^i_{\mu\nu} T^i + (1-P) \eta^3_{\mu\nu} \bigr).
\end{equation}
It is then straightforward to derive the fancy formula (\ref{nc-curved}) using Eqs. (\ref{vierbein-1})
and (\ref{f-proof}).

One can similarly understand the self-duality equation (\ref{curved-sde2}). Let us define
\begin{equation}\label{b-form}
B= \frac{1}{2} B_{ab} E^a \wedge E^b = - \frac{1}{\theta} \eta^3_{ab} E^a \wedge E^b
\equiv - \frac{2}{\theta} \Omega
\end{equation}
where we used the relation $B_{ab} = - \frac{2}{\theta} \eta^3_{ab}$. Then the self-duality
equation (\ref{curved-sde2}) is automatically satisfied since Eq. (\ref{b-form}) can be written
in the form (\ref{curved-sde2}) (with $+$-sign). That is, if we understand the background $B$-field
as $B_{\mu\nu} \equiv - \frac{2}{\theta}  E^{~a}_\mu \eta^3_{ab} E^b_{~\nu}$,
a straightforward calculation shows that
\begin{eqnarray}\label{kahler-b}
    B_{\mu\nu}(x) &=&  2\bigl( - P \eta^3_{\mu\nu} + Q \overline{\eta}^i_{\mu\nu} T^i \bigr)
    = -2 \bigl(G \eta^3 \bigr)_{\mu\nu}(x) \nonumber \\
    &=&- 2 \eta^3_{\mu\nu} +  F_{\mu\nu}(x)
\end{eqnarray}
where we used Eq. (\ref{f-proof}). It is obvious that $B = \frac{1}{2} B_{\mu\nu}(x) dx^\mu \wedge dx^\nu$
is a closed two-form, i.e. $dB=0$ as long as $dF=0$--the Bianchi identity.
Then Eq. (\ref{b-form}) implies that $\Omega$ is the K\"ahler form of the metric (\ref{eff-vierbein})
and (1,1)-form with respect to the complex structure ${J^a}_b = \eta^3_{ab}$ \cite{prl-06,hsy-epl09}.
In the end, we got a very nice interpretation of the self-duality equations
(\ref{curved-sde}) and (\ref{curved-sde2}) consistent with general relativity.

It is straightforward to generalize the bottom-up approach to a space with the metric (\ref{metric-form}).
After a little algebra we find that the metric (\ref{metric-form}) can be written as
\begin{equation}\label{gmetric-gen}
g_{\mu\nu}(x) = \frac{1}{2}(A^2 + B^2) \delta_{\mu\nu} - \frac{1}{r^2} (A^2-B^2)
(\eta^3 \overline{\eta}^i)_{\mu\nu} T^i
\end{equation}
and so the $U(1)$ field strength in Eq. (\ref{g-f}) is given by
\begin{equation}\label{u1f-gen}
F_{\mu\nu}(x) = f_1(r) \eta^3_{\mu\nu} + f_2(r) \overline{\eta}^i_{\mu\nu} T^i
\end{equation}
where
\begin{equation}\label{ab-metric}
    f_1(r) = 1 - \frac{1}{2}(A^2 + B^2), \qquad f_2(r) = - \frac{1}{r^2} (A^2-B^2).
\end{equation}
Using the result (\ref{u1f-gen}) one can also calculate the inverse metric
\begin{eqnarray}\label{inv-G}
    G^{\mu\nu} &=& \Bigl( \frac{1}{1 + F\theta} \Bigr)^{\mu\nu} \nonumber \\
    &=& \frac{2+A^2+B^2}{(1+A^2)(1+B^2)} \delta_{\mu\nu}
    + \frac{2(A^2-B^2)}{r^2(1+A^2)(1+B^2)} \bigl(\eta^3 \overline{\eta}^i \bigr)_{\mu\nu} T^i
\end{eqnarray}
and the field strength (\ref{symp-f}) of symplectic $U(1)$ gauge fields
\begin{equation}\label{u1f-ncgen}
\widehat{F}_{\mu\nu}(x) = \frac{2}{(1+A^2)(1+B^2)} \Bigl[ (1-A^2B^2) \eta^3_{\mu\nu}
-\frac{2}{r^2} (A^2-B^2) \overline{\eta}^i_{\mu\nu} T^i \Bigr].
\end{equation}
It is easy to check the result in Refs. \cite{sty-06,prl-06,hsy-epl09} that
symplectic $U(1)$ instantons are equivalent to gravitational instantons.
First note that the $U(1)$ field strength (\ref{u1f-ncgen}) with $A^2B^2 = 1$ obeys the self-duality
equation (\ref{self-dual-inst}). In this case the spin connections in Eq. (\ref{spin-conn})
become anti-self-dual, i.e., $\omega_{ab} + \frac{1}{2} {\varepsilon_{ab}}^{cd} \omega_{cd} = 0$,
which leads to curvature tensors satisfying $R_{ab} + \frac{1}{2} {\varepsilon_{ab}}^{cd}
R_{cd} = 0$. For example, the Eguchi-Hanson metric (\ref{eh-metric}) clearly satisfies
$A^2B^2 = 1$ for which the $U(1)$ field strength (\ref{u1f-ncgen}) becomes anti-self-dual.
Therefore the metric (\ref{gmetric-gen}) with $A^2B^2 = 1$ is a gravitational instanton.

A geometrical meaning of the 't Hooft symbols defined in Appendix A
is to specify the triple $(I,J,K)$ of complex structures of
$\mathbb{R}^4$ for a given orientation. Therefore the choice of a
particular NC parameter in Eq. (\ref{hodge-theta}), e.g.
$\theta^{\mu\nu} = \frac{\theta}{2} \eta^3_{\mu\nu}$, corresponds to
singling out a particular complex structure, for example, $J =
T^3_+$ in Eq. (\ref{t+}). And the space (\ref{metric-form}) inherits
the complex structure $J$ from $\mathbb{R}^4$. So let us consider
the fundamental 2-form defined by
\begin{equation}\label{kahler-form}
    \omega = \frac{1}{2} \eta^3_{ab} e^a \wedge e^b.
\end{equation}
Now we will show that the fundamental 2-form $\omega$ is closed,
i.e. $d\omega = 0$, and so defines the K\"ahler form of the metric
(\ref{metric-form}) as far as the $U(1)$ field strength
(\ref{u1f-gen}) obeys the Bianchi identity, i.e. $dF=0$. In other
words, the metric (\ref{metric-form}) is always K\"ahler if and only
if  $dF=0$. Using the covectors in Eq. (\ref{co-vector}) and
coframes in Eq. (\ref{coframe-s3}), the 2-form $\omega$ can be
written as
\begin{equation}\label{kahler-cart}
    \omega = \frac{1}{2} \Bigl[ \frac{1}{2} (A^2 + B^2) \eta^3_{\mu\nu} +
    \frac{1}{r^2} (A^2 - B^2) \overline{\eta}^i_{\mu\nu} T^i \Bigr] dx^\mu \wedge dx^\nu
\end{equation}
where we used the identity (\ref{id-bn}) and (\ref{eta-o4-algebra}).
After using the result (\ref{u1f-gen}), the 2-form in Eq. (\ref{kahler-cart}) finally reduces to
\begin{equation}\label{kahler-fin}
    \omega = \frac{1}{2} \eta^3_{\mu\nu} dx^\mu \wedge dx^\nu -
    \frac{1}{2} F_{\mu\nu} dx^\mu \wedge dx^\nu = \omega^{(0)} - F
\end{equation}
where $\omega^{(0)} = \frac{1}{2} \eta^3_{\mu\nu} dx^\mu \wedge dx^\nu$.
Consequently, $d\omega = 0$ if and only if $dF=0$.\footnote{One can easily see that the metric
(\ref{gmetric-gen}) can be written as $g_{\mu\nu}(x) = - \omega_{\mu\lambda}(x) \eta^3_{\lambda\nu}$
where $\omega_{\mu\nu}(x)$ is defined by Eq. (\ref{kahler-cart}).
This is nothing but the definition of K\"ahler form with the complex structure ${J^\mu}_\nu
= \eta^3_{\mu\nu}$, i.e., $\omega(X,Y) = g(X,JY)$ for vector fields $X,Y \in TM$.}
For the $U(1)$ field strength (\ref{u1f-gen}), the Bianchi identity, $dF=0$, is reduced to the
first order differential equation given by
\begin{equation}\label{bi-kahler}
    \frac{dB^2}{dr} = \frac{2}{r}(A^2-B^2).
\end{equation}

One may wonder whether one can get $U(1)$ gauge fields in the same way from the Taub-NUT metric
\cite{egh-report} which takes the form
\begin{equation}\label{taub-nut}
    ds^2 = \frac{1}{4} \frac{\rho + m}{\rho -m} d\rho^2 + (\rho^2-m^2) (\sigma_1^2 + \sigma_2^2)
    + 4m^2 \frac{\rho -m}{\rho + m} \sigma_3^2.
\end{equation}
A critical difference from the Eguchi-Hanson metric
(\ref{eh-metric}) is that the Taub-NUT metric (\ref{taub-nut}) is
locally asymptotic at infinity to $\mathbb{R}^3 \times \mathbb{S}^1$
and so it belongs to the class of asymptotically locally flat (ALF)
spaces. Therefore the Taub-NUT metric cannot be represented by the
Hopf coordinates (\ref{hopf}) and it is difficult to naively
generalize the previous construction to ALF spaces. From the gauge
theory point of view, it may be related to the fact that ALF spaces
arise from NC monopoles \cite{lee-yi} whose underlying equation is
defined by an $\mathbb{S}^1$-compactification of Eq.
(\ref{self-dual-inst}), the so-called Nahm equation. We will discuss
in Ref. \cite{future-paper} a possible generalization to include the
Taub-NUT metric (\ref{taub-nut}) in the bottom-up approach of
emergent gravity.

\subsection{Gravitational metric from Braden-Nekrasov instanton}

Braden and Nekrasov \cite{bn-inst} considered a deformed Atiyah-Drinfeld-Hitchin-Manin (ADHM) construction\footnote{\label{adhm-comp}Here
we refer the deformed ADHM construction to $\tau_z \tau_z^\dagger - \sigma_z^\dagger \sigma_z
= 2 \zeta_{\mathbb{R}}, \; \tau_z \sigma_z = \zeta_{\mathbb{C}}$ whereas the undeformed ADHM construction
to $\tau_z \tau_z^\dagger - \sigma_z^\dagger \sigma_z = 0, \; \tau_z \sigma_z = 0$
where $\zeta_{\mathbb{R}} = \zeta^3$ and $\zeta_{\mathbb{C}} = \zeta^1 + i \zeta^2$.} on {\it commutative}
$\mathbb{C}^2$ whose solution gives a resolved moduli space $\widetilde{\mathcal{M}}_{N,k}
= \mu^{-1}(\vec{\zeta})/U(k)$ in terms of hyper-K\"ahler quotient where
$\mu^{-1}(\vec{\zeta})$ are $U(k)$ hyper-K\"ahler moment maps \cite{inst-review1,inst-review2}.
It was shown in Ref. \cite{ns-inst} that the same resolved instanton moduli space arises by considering
the standard (undeformed) ADHM construction but instead assuming the spacetime coordinates
having the commutation relation (\ref{nc-space}). In this case the deformation parameters $\vec{\zeta}$
in the $U(k)$ hyper-K\"ahler moment maps $\mu^{-1}(\vec{\zeta})$ arise from $\theta^{(\pm)i}$
in Eq. (\ref{hodge-theta}).
Thus it will be interesting to study the relation between the deformed ADHM construction of
an ordinary commutative gauge theory and the undeformed ADHM construction of a NC gauge theory.
Furthermore, as we discussed in Sec. I, the NC gauge theory can be mapped to the ordinary
commutative gauge theory by the SW map. Therefore one may expect that NC $U(1)$ instantons
constructed in Ref. \cite{ns-inst} would be related by the SW map to $U(1)$ instantons
constructed by the deformed ADHM data on commutative $\mathbb{C}^2$. Interestingly, as already noted
in Ref. \cite{bn-inst} (see {\it Notes added five years later} in section 6.2),
the commutative $U(1)$ instantons are not related to the NC $U(1)$ instantons by the SW map
and the commutative description of the Nekrasov-Schwarz (NS) instanton is indeed different from
the Braden-Nekrasov (BN) instanton as was shown in Ref. \cite{ks-inst02}.

We want to shed light on this puzzle by studying the geometrical properties of a four-manifold
determined by the BN $U(1)$ instanton in the context of emergent gravity.
We will compare the result with the NS instanton whose SW map gives rise to
a complete regular geometry described by the Eguchi-Hanson metric as was shown in Sec. III.A.
This analysis reveals that the NC structure of spacetime is essential for
a smooth topology change of spacetime and a resolution of spacetime singularities
\cite{our-topch}. So it will be interesting to consider a full NC deformation for a complete explanation.
A smooth topology change was also addressed in Ref. \cite{topch1,topch2} for two-dimensional NC Riemann
surfaces. See also a recent review in Ref. \cite{martinec} for the significance of NC geometry
for the topology change and singularity resolution in string theory.

It turns out \cite{bn-inst} that $U(1)$ instantons constructed from the deformed ADHM construction
on commutative $\mathbb{C}^2$ are still singular unlike NC $U(1)$ instantons
and so it is necessary to change the topology
of spacetime in order to make the corresponding $U(1)$ gauge fields non-singular. The reason that
the Abelian instanton exists is that spacetime is now blown up and there are non-contractible 2-spheres.
Then the resulting spacetime is not $\mathbb{C}^2$ but a K\"ahler manifold $X$ which is a blowup of
$\mathbb{C}^2$ at a finite number of points. In the end $U(1)$ gauge fields on $X$ are well-defined
and carry a nontrivial first Chern class when the gauge fields are restricted to exceptional divisors
in addition to a nontrivial second Chern class $k$. But the blowup becomes
manifest only by gluing local coordinate patches and performing a proper gauge transformation
on their intersections. For example, one can choose local coordinates $(t, \lambda)$ on a patch
$\mathcal{U}_0$ such that $z_1 = t, \; z_2 = t \lambda$ where $(z_1, z_2) \in \mathbb{C}^2$
and another local coordinates $(s, \mu)$ on another patch $\mathcal{U}_\infty$
such that $z_1 = \mu s, \; z_2 = s$. On these patches $\mathcal{U}_0$ and $\mathcal{U}_\infty$,
the point $0 = (0,0)$ in $\mathbb{C}^2$ is replaced by the space $\mathbb{C}\mathbb{P}^1$ of
complex lines passing through the point $0$. One can use the local coordinates to represent $U(1)$
gauge fields on each local chart and then extend them via a gauge transformation
to a safe region where $(z_1, z_2) \neq 0$ \cite{bn-inst}.

Now let us undertake a more systematic investigation of the single $U(1)$ instanton
in Sec. 4 of Ref. \cite{bn-inst}. The instanton gauge fields are given by (setting $\theta = 1$)
\begin{equation}\label{bna-sol}
    A = \frac{1}{2r^2(1+r^2)}(z_1 d\overline{z}_1 - \overline{z}_1 dz_1
    + z_2 d\overline{z}_2 - \overline{z}_2 dz_2)
\end{equation}
and
\begin{equation}\label{bnf-sol}
    F = \frac{dz_1 \wedge d\overline{z}_1 + dz_2 \wedge d\overline{z}_2}{r^2(1+ r^2)}
    - \frac{1 + 2r^2}{r^4(1+r^2)^2}\sum_{i,j} z_i \overline{z}_j dz_j \wedge
    d\overline{z}_i
\end{equation}
where $z_1 = x^1 + ix^2, \; z_2 = x^3 + i x^4$ and $r^2 = |z_1|^2 + |z_2|^2$.
The above gauge fields $A \equiv 2i A_\mu dx^\mu$ (the factor 2 scaling is
just for convenience) and $F = dA = i F_{\mu\nu} dx^\mu \wedge dx^\nu$ can be represented
in terms of the Cartesian coordinates $\{x^\mu\}$ and can be written as
\begin{equation}\label{bna-c}
    A_\mu(x) = \frac{t^2}{2r^2 (t^2 + r^2)} \eta^3_{\mu\nu} x^\nu
\end{equation}
and
\begin{equation}\label{bnf-c}
    F_{\mu\nu} (x) = - \frac{t^4}{2r^2 (t^2 + r^2)^2} \eta^3_{\mu\nu}
    + \frac{t^2(t^2+2r^2)}{r^4 (t^2 + r^2)^2}
    \overline{\eta}^i_{\mu\nu} T^i,
\end{equation}
where the dimensionful parameter $t^2 = \theta$ may be useful for an accessible comparison
with the NS instanton.
To get the expression (\ref{bnf-c}), we used the identity (\ref{id-bn}).
It may be interesting to compare the asymptotic behaviors (set $t=1$) of the NS instanton (\ref{u1-eh})
and the BN instanton (\ref{bnf-c}) both in $r \to \infty$
\begin{eqnarray} \label{asymp-ns}
&&  \mathrm{NS}: \; F_{\mu\nu}(x) = \frac{1}{r^6} \Bigl(1-\frac{1}{2r^4} + \cdots \Bigr)
\overline{\eta}^i_{\mu\nu} T^i
  - \frac{1}{8r^8} \Bigl(1-\frac{1}{r^4} + \cdots \Bigr) \eta^3_{\mu\nu}, \nonumber \\
&&  \mathrm{BN}: \; F_{\mu\nu}(x) = \frac{2}{r^6} \Bigl(1-\frac{3}{2r^2} + \cdots \Bigr)
\overline{\eta}^i_{\mu\nu} T^i
  - \frac{1}{2r^6} \Bigl(1-\frac{2}{r^2} + \cdots \Bigr) \eta^3_{\mu\nu},
\end{eqnarray}
and in $r \to 0$
\begin{eqnarray} \label{asymp-bn}
&& \mathrm{NS}: \; F_{\mu\nu}(x) = \frac{1}{r^4} \Bigl(1-\frac{1}{2}r^4 + \frac{3}{8}r^8 + \cdots \Bigr) \overline{\eta}^i_{\mu\nu} T^i
  - \frac{1}{2r^2} \Bigl(1-2r^2 + \frac{3}{2} r^4 + \cdots \Bigr) \eta^3_{\mu\nu}, \nonumber \\
&& \mathrm{BN}: \; F_{\mu\nu}(x) = \frac{1}{r^4} \Bigl(1-r^4 + 2r^6 + \cdots \Bigr)
\overline{\eta}^i_{\mu\nu} T^i
  - \frac{1}{2r^2} \Bigl(1- 2r^2 + 3r^4 + \cdots \Bigr) \eta^3_{\mu\nu}.
\end{eqnarray}
One can see \cite{bn-inst} that the asymptotic behaviors for two instantons are almost the same except that
the BN instanton is slightly slowly decaying at $r \to \infty$.

Note that the instanton gauge field (\ref{bna-c}) was obtained through the ADHM construction.
Nevertheless, as was observed in Eq. (3.14) in Ref. \cite{bn-inst}, the completeness relation
of ADHM data fails at a finite number of points which brings about the field strength (\ref{bnf-c})
being neither self-dual nor anti-self-dual. A notable point is that the commutative description
of the NS instantons also shares this feature as shown in Eq. (\ref{u1-eh}).
But Eq. (\ref{esw-nsu1}) verifies that the NS instanton becomes (anti-)self-dual
in the NC description. Thus one may wonder whether the same property can be realized even for the BN instanton. To see what happens in the NC description of the BN instanton, let us apply the SW map (\ref{eswmap})
to the $U(1)$ field strength (\ref{bnf-c}). The result is given by Eq. (\ref{u1f-ncgen}) with
\begin{equation}\label{bn-ab}
    A^2(r) = \frac{r^2(r^2 + 2t^2)}{(r^2+t^2)^2}, \qquad B^2(r) = \frac{r^4 + r^2t^2 + t^4}{r^2(r^2+t^2)}.
\end{equation}
Explicitly it takes the form
\begin{equation}\label{bnf-nc}
    \widehat{F}_{\mu\nu} (x) = \frac{2t^2}{(2r^4+2r^2t^2 + t^4)(2r^4+4r^2t^2+t^4)}
    \Bigl[ \frac{2(r^2+t^2)(2r^2 + t^2)}{r^2}
    \overline{\eta}^i_{\mu\nu} T^i  - r^2 t^2 \eta^3_{\mu\nu} \Bigr].
\end{equation}
The result (\ref{bnf-nc}) shows that the BN instanton is neither self-dual nor anti-self-dual
even in the NC description. It can be understood as follows. First note that the field strength (\ref{bnf-c})
takes the form (\ref{u1f-gen}) and the resultant NC field strength is then given by Eq. (\ref{u1f-ncgen})
where $A$ and $B$ are given by (\ref{bn-ab}).
But the coefficients in Eq. (\ref{bnf-c}) does not satisfy the relation $A^2B^2 = 1$,
which leads to the former conclusion.
This presents a sharp contrast with the NS instantons with (anti-)self-dual curvatures in NC spacetime.

It may be interesting to check whether the identity (\ref{nc-curved}) is true for the BN instanton.
That is, one can ask whether the NC field strength (\ref{bnf-nc}) can be written as $\widehat{F}_{ab}
= E_a^{~\mu} F_{\mu\nu} E^\nu_{~b}$ with the commutative field strength (\ref{bnf-c})
where the vierbeins are defined by $G_{\mu\nu} =
\delta_{\mu\nu} + (F\theta)_{\mu\nu} = E_\mu^a E_\nu^a$.
We checked that the identity (\ref{nc-curved}) still holds for the BN instanton though the field strength
is neither self-dual nor anti-self-dual. The proof goes through the same way as the NS instanton case.
We will present in Sec. V a proof of the identity (\ref{nc-curved}) for general $U(1)$ gauge fields
with the symmetric metric (\ref{eff-metric}).

We analyzed before the asymptotic behavior of the BN instanton and found that the leading behavior is
exactly the same as the NS instanton. An interesting question is then whether a four-dimensional manifold
determined by the BN instanton also exhibits a similar geometrical countenance.
In order to investigate the geometrical properties of the four-manifold, let us consider
the metric $ds^2 = g_{\mu\nu}(x) dx^\mu dx^\nu = e^a \otimes e^a$ defined by Eq. (\ref{g-f})
with the solution (\ref{bnf-c}). But we want to express the metric in the form (\ref{metric-form})
using the left-invariant one-forms in Eq. (\ref{coframe-s3}). To implement this form,
one can employ the reverse procedure of Sec. III.A to arrive at the result
\begin{equation}\label{bn-metric}
ds^2 = \frac{r^2(r^2 + 2t^2)}{(r^2+t^2)^2} (dr^2 + r^2 \sigma_3^2)
+ \frac{r^4 + r^2t^2 + t^4}{r^2+t^2} (\sigma_1^2 + \sigma_2^2)
\end{equation}
and so $A^2$ and $B^2$ are given by Eq. (\ref{bn-ab}). This metric form would indicate
that the four-manifold described by Eq. (\ref{bn-metric}) might be akin to
the Eguchi-Hanson metric (\ref{eh-metric}). For example, the metric (\ref{bn-metric}) also contains
a nontrivial two-cycle $\mathbb{S}^2$ at the origin $(r=0)$ where the metric is degenerated to the two
dimensional sphere with the metric $t^2(\sigma_1^2 + \sigma^2_2)$. The identification of topological
invariants discussed in Sec. IV indicates that the two-sphere in Eq. (\ref{bn-metric}) is related to the K\"ahler blowup for the BN instanton at the origin of $\mathbb{C}^2$.
In order to understand the engrossing feature, let us recapitulate a corresponding aspect
for the NS instanton \cite{hsy-ijmp09}. The Eguchi-Hanson metric (\ref{eh-metric}) has a curvature
that reaches a maximum at the `origin' $\rho = t$ (recall that $\rho^4 = r^4 + t^4$),
falling away to zero in all four directions as the radius $\rho$ increases.
Since the radial coordinate runs down only as far as $\rho = t$, there is a minimal 2-sphere $\mathbb{S}^2$
of radius $t$ described by the metric $t^2(\sigma_1^2 + \sigma_2^2)$. This degeneration of the generic
three dimensional orbits to the two dimensional sphere is known as a `bolt' \cite{bolt}.
But, $\rho = t$ corresponds to the origin $r = 0$ of the embedding coordinates in field theory and so
this nontrivial topology is not visible in the gauge theory description.
However, as we showed in section 3.1, the emergent gravity approach where a Riemannian manifold
is emerging from dynamical gauge fields reveals a nontrivial topology of NC $U(1)$ gauge fields.
As one can see from Eq. (\ref{g-f}),
if $F=0$, the corresponding spacetime becomes $\mathbb{R}^4$ without any nontrivial cycles
but, if the intanton gauge fields in Eq. (\ref{u1-eh}) are developed, the spacetime evolves
to the Eguchi-Hanson space which contains a noncontractible 2-sphere dubbed as the bolt.
Therefore the emergent gravity verifies the topology change of spacetime due to $U(1)$ instantons.
Exactly the same phenomenon happened for the BN instanton.
But more detailed analysis brings some surprise.

It is straightforward to calculate the spin connections and curvature tensors of the metric (\ref{bn-metric})
using the results of Appendix B. We present the explicit result for reader's convenience:
\begin{eqnarray} \label{bn-spin}
   && \omega_{12} = - \frac{r^6 + 2r^4t^2 + 4r^2t^4 + 2t^6}{r^2\sqrt{r^2+2t^2}(r^4+r^2t^2+t^4)} e^3,
   \qquad \omega_{34} = \frac{r^4 + 3r^2t^2 + 4t^4}{r^2(r^2+ 2t^2)^{3/2}} e^3,  \nonumber \\
  &&  \omega_{13} = - \omega_{42} =  \frac{r^2\sqrt{r^2+2t^2}}{r^4+r^2t^2+t^4} e^2,
  \qquad \omega_{14} = - \omega_{23} =  \frac{r^2\sqrt{r^2+2t^2}}{r^4+r^2t^2+t^4} e^1,
\end{eqnarray}
and
\begin{eqnarray} \label{bn-curvature}
   && R_{12} = X e^1 \wedge e^2 - 2 Y e^3 \wedge e^4,
   \qquad \qquad R_{34} = - 2Y e^1 \wedge e^2 + Z e^3 \wedge e^4,  \nonumber \\
  &&  R_{13} = - R_{42} =  Y (e^3 \wedge e^1 - e^2 \wedge e^4) ,
  \quad R_{14} = - R_{23} =  Y (e^2 \wedge e^3 - e^1 \wedge e^4),
\end{eqnarray}
where the vierbeins $e^a$ are given by Eq. (\ref{co-vector}) with the result (\ref{bn-ab})
and $X,Y,Z$ are given by
\begin{eqnarray} \label{xyz}
  && X = \frac{4t^4(2r^2+t^2)}{(r^4 + r^2t^2 + t^4)^2}, \nonumber \\
  && Y = \frac{t^4(4r^4 + 7r^2t^2 + 4t^4)}{(r^2+ 2t^2)(r^4 + r^2t^2 + t^4)^2}, \\
  && Z = \frac{4t^4(2r^2+3t^2)}{r^2(r^2 + 2 t^2)^3}. \nonumber
\end{eqnarray}
Using the result (\ref{bn-curvature}), one can easily read off the Ricci tensor $R_{ab} \equiv
R_{acbc}$ and the Ricci scalar $R \equiv R_{aa}$. We present only the diagonal Ricci tensors
which read as
\begin{eqnarray} \label{bn-ricci}
   && R_{11} = R_{22} = \frac{6r^2 t^6}{(r^2 + 2t^2)(r^4 + r^2 t^2 + t^4)^2},  \nonumber \\
   && R_{33} = R_{44} = - \frac{6t^6 (3r^8 + 8r^6 t^2 + 6r^4 t^4 -2t^8)}{r^2(r^2+ 2t^2)^3(r^4 + r^2t^2 + t^4)^2}.
\end{eqnarray}
Therefore the Ricci scalar is given by
\begin{equation}\label{bn-scalar}
    R = - \frac{24 t^6 (r^4 + r^2 t^2 - t^4)}{r^2(r^2+ 2t^2)^3(r^4 + r^2t^2 + t^4)}.
\end{equation}

Some remarks are in order. We proved in Eq. (\ref{kahler-fin}) that the metric (\ref{metric-form})
becomes K\"ahler if the $U(1)$ field strength (\ref{u1f-gen}) satisfies
the Bianchi identity $dF=0$. Note that the metric (\ref{bn-metric}) was derived from the
$U(1)$ field strength (\ref{bnf-sol}) which satisfies the Bianchi identity $dF=0$. Therefore
the metric (\ref{bn-metric}) must be K\"ahler. Indeed it is easy to check that
the coefficients in Eq. (\ref{bn-ab}) obeys the differential equation (\ref{bi-kahler}).
If one looks at the curvature tensors in Eq. (\ref{bn-curvature}),
one can see that the four-manifold generated by the BN instanton is neither self-dual
nor anti-self-dual though it is close to an anti-self-dual manifold. This is consistent with
the gauge theory result. Moreover it has a nontrivial Ricci scalar which is divergent at the origin.
This indicates that the classical geometry emergent from the BN instanton contains a spacetime
singularity at the origin. To see that this is a true singularity, one must look at quantities
that are independent of the choice of coordinates. An obvious candidate is of course the Ricci scalar
$R$ which is already singular in our case. But it may identically vanish for some solutions,
for example, the Schwarzschild black hole. For such cases, another important quantity is the Kretschmann
invariant which is defined by $K = R_{\mu\nu\rho\sigma}R^{\mu\nu\rho\sigma}$.
The existence of the singularity can be verified by noting that either the Ricci scalar $R$ or
the Kretschmann scalar $K$ is infinite. For example, the famous Schwarzschild black hole exhibits
such a spacetime singularity for which the Kretschmann scalar $K$ is given by
\begin{equation} \label{st-singularity}
   K = R_{\mu\nu\rho\sigma} R^{\mu\nu\rho\sigma} = \frac{48G^2 M^2}{r^6}
\end{equation}
which blows up at $r=0$ indicating the presence of a spacetime singularity. Thus one can calculate
the Kretschmann scalar $K$ for the metric (\ref{bn-metric}) in order to further confirm
the spacetime singularity and the result is given by
\begin{equation}\label{bn-k}
   \frac{K}{64 t^8} =  \frac{(2r^2 + 3 t^2)^2}{r^4(r^2+ 2t^2)^6} +
   \frac{(2r^2 + t^2)^2}{(r^4 + r^2t^2 + t^4)^4} +
   \frac{(4r^4 + 7 r^2 t^2 + 4 t^4)^2}{(r^2+ 2t^2)^2(r^4 + r^2t^2 + t^4)^4}.
\end{equation}
One can clearly see that the first term becomes divergent at the origin although
the next two terms are regular everywhere as long as $t^2 = \theta \neq 0$.
And so Eq. (\ref{bn-k}) again verifies the spacetime singularity at the origin.

Let us try to understand why the classical geometries generated by the BN instanton and the NS instanton
are so dissimilar, especially, in view of the singularity structure.
From the gauge theory point of view, both instantons are constructed by solving
the ADHM data for the rank one gauge group. The hyper-K\"ahler moment maps $\mu^{-1}(\vec{\zeta})$
are deformed, i.e. $\vec{\zeta} \neq 0$, for both cases.
But the origin is different. (See the footnote \ref{adhm-comp}.) For the BN instanton case,
the ADHM data are defined on {\it commutative} $\mathbb{C}^2$ and the non-vanishing deformation parameters $\vec{\zeta}$ are assumed from the beginning. But, for the NS instanton case,
the non-vanishing deformation parameters $\vec{\zeta}$ are not introduced by hand.
Instead the ADHM data are defined on {\it noncommutative} $\mathbb{C}^2$ obeying the Heisenberg algebra
\begin{equation}\label{h-algebra}
    [z_i, z_j^\dagger] = \zeta_\mathbb{R} \delta_{ij}, \qquad i,j =1,2,
\end{equation}
where $\zeta_\mathbb{R} = \frac{1}{2} \eta^3_{\mu\nu}\theta^{\mu\nu}$.
The deformation parameter $\zeta_\mathbb{R}$ appears in the moment maps $\mu^{-1}(\vec{\zeta})$ due to
the NC algebra (\ref{h-algebra}). Furthermore the NC space (\ref{h-algebra}) resolves the singularities
of instanton moduli space coming from point-like instantons which shrink to zero size \cite{ns-inst}.
Therefore the NC instantons are well-defined
and nonsingular. However, this feature is lacking if the deformed ADHM data are defined over
a commutative space. The completeness relation fails at a finite number of points, called ``freckles."
As a result, the ADHM gauge fields are no longer (anti-)self-dual as one can see from Eq. (\ref{bnf-c})
(see also Eq. (3.14) in Ref. \cite{bn-inst}). The spacetime singularity in Eqs. (\ref{bn-scalar}) and (\ref{bn-k}) arises at the freckle where the instanton is placed.
This argument implies  \cite{our-topch} that, in order to describe the topology change
of spacetime and the resolution of spacetime singularity,
it is not enough to deform only the ADHM data leaving spacetime to be commutative.
The NC structure of spacetime is essential to resolve the spacetime singularities in general relativity.
Consequently it is necessary to take the NC space (\ref{nc-space}) at the outset and then consider
the commutative description of NC gauge theory.

\section{Topological invariants of $U(1)$ gauge fields}

The emergent gravity raises an intriguing question about topological invariants in gravity and
$U(1)$ gauge theory. In the gravity side, there are two topological invariants associated with
the Atiyah-Patodi-Singer index theorem for an elliptic complex in four dimensions \cite{egh-report,besse},
namely the Euler characteristic $\chi(M)$ and the Hirzebruch signature $\tau(M)$,
which can be expressed as integrals of the curvature of a four-manifold.
But there is no natural topological invariant in ordinary $U(1)$ gauge theory.
For example, the second Chern class of $U(1)$ bundle
on $\mathbb{R}^4$ is trivial and it is easy to show that a nontrivial instanton charge is
incompatible with the vanishing of $F=dA$ at infinity.
The second Chern class of $U(1)$ bundle can be well-defined only for NC $U(1)$ instantons
\cite{ns-inst,nek-cmp03}. An interesting point is that the commutative limit of
NC $U(1)$ instantons is equivalent to gravitational instantons \cite{hsy-jhep09,hsy-epl09}.
Hence the emergent gravity implies that the commutative limit of NC $U(1)$ gauge fields has to carry
the same topological invariants as four-dimensional Riemannian manifolds.
Thus a natural question is how to construct the two topological invariants of four-manifolds
in terms of $U(1)$ gauge fields in the context of emergent gravity.

First it will be interesting to compare the instanton number for the NS and BN instantons.
Using the results (\ref{u1-eh}) and (\ref{bnf-c}), one can get
\begin{eqnarray} \label{nsinst-den}
\mathrm{NS}: && F \wedge F =  - \frac{2}{r^2\sqrt{r^4 + t^4}} \bigl(\sqrt{r^4 + t^4} - r^2 \bigr)^2 d^4 x, \\
\label{bninst-den}
\mathrm{BN}: && F \wedge F = - \frac{8t^2}{r^2 (r^2 + t^2)^3} d^4 x
\end{eqnarray}
and so the instanton number is given by\footnote{Here we are considering the anti-self-dual instantons with
real $F$ and adopt the normalization $1/4\pi^2$ in Ref. \cite{bn-inst}
for the instanton number $I$ which is different from $1/8 \pi^2$ in Eq. (\ref{ncinst-num}).
Since $F \wedge F = d(A \wedge F)$ and $F \to 0$ as $r \to \infty$, it is obvious that the contribution
to the instanton number $I$ is localized at the origin.}
\begin{eqnarray} \label{nsinst-num}
\mathrm{NS}: && I = \frac{1}{4\pi^2 } \int F \wedge F =  - \frac{t^4}{4}, \\
\label{bninst-num}
\mathrm{BN}: && I = \frac{1}{4\pi^2 } \int F \wedge F =  -1.
\end{eqnarray}
It is amusing to notice that the instanton number for the NS instanton depends on $t^4 = \theta^2$
while the BN instanton does not. Actually this fact for the former case was observed
in Ref. \cite{ncft-sw} and was interpreted as a nonperturbative breakdown of the SW map due
to a finite radius of convergence. But the BN instanton
solution (\ref{bna-sol}) was directly obtained by solving the ADHM equations where the instanton
number $k = |I|$ specifies the dimension of the vector space $\mathbb{C}^{N+2k}$. Thus
$k$ should be an integer number for consistency. For the same reason, the instanton number
for NC $U(1)$ instantons satisfying the self-duality equation (\ref{self-dual-inst}) must be an integer
$k \in \mathbb{Z}$ \cite{chu-npb02,kly-jkps02,kly-plb01} which is defined by
\begin{equation} \label{ncinst-num}
I = \frac{1}{8\pi^2 } \int \widehat{F} \wedge \widehat{F} \in \mathbb{Z}
\end{equation}
where $\widehat{F} = d\widehat{A} - i \widehat{A} \wedge \widehat{A}$.
The commutative description of NS instantons is rather puzzling because the instanton number is not quantized
and so the topology of $U(1)$ gauge fields becomes obscure. But, as we argued above, the commutative limit
of NC $U(1)$ gauge fields carries the topological information in the form of four-dimensional
Riemannian manifolds. In a deep NC space where the continuum description in terms of smooth geometries
becomes bad, the NC $U(1)$ gauge bundle whose invariant is given by Eq. (\ref{ncinst-num}) will take over
the topological information.

We will now investigate with explicit examples in Sec. III how the topological information of
$U(1)$ gauge fields is reflected in a four-dimensional Riemannian manifold.
The topological invariants for four-manifolds have a local expression due to the Atiyah-Singer
index theorem \cite{egh-report,besse}. For a general Riemannian manifold $M$, the Euler number $\chi(M)$
for the de Rham complex and the signature $\tau(M)$ for the Hirzebruch signature complex are defined by
\begin{eqnarray} \label{euler}
\chi(M) & = & \frac{1}{32 \pi^2} \int_{M }\varepsilon^{abcd}
R_{ab} \wedge R_{cd} + \frac{1}{16 \pi^2} \int_{\partial M}
\varepsilon^{abcd} \Big( v_{ab} \wedge R_{cd}
- \frac{2}{3} v_{ab} \wedge v_{ce} \wedge v_{ed} \Big), \\
\label{signature}
\tau(M) &=& - \frac{1}{24 \pi^2} \int_{M} \mathrm{Tr} R \wedge R
 - \frac{1}{24 \pi^2} \int_{\partial M}
\mathrm{Tr} v \wedge R + \eta_S(\partial M),
\end{eqnarray}
where $v_{ab}$ is the second fundamental form of the boundary
$\partial M$. It is defined by
\begin{equation} \label{2-f-form}
v_{ab} = \omega_{ab} - \omega_{0ab},
\end{equation}
where $\omega_{ab}$ are the actual connection 1-forms and $\omega_{0ab}$ are the connection 1-forms
if the metric were locally a product form near the boundary \cite{egh-report}.
The connection 1-form $\omega_{0ab}$ has only tangential
components on $\partial M$ and so the second fundamental form
$v_{ab}$ has only normal components on $\partial M$.
And $\eta_S(\partial M)$ is the $\eta$-function given by the eigenvalues
of a signature operator defined over $\partial M$ and depends only
on the metric on $\partial M$ \cite{egh-report}. The topological invariants are also related to
nuts (isolated points) and bolts (two surfaces), which are the fixed points of the action of
one parameter isometry groups of gravitational instantons \cite{bolt}.
Using the gauge theory formulation that Einstein gravity can be formulated
as a gauge theory of Lorentz group $SO(4) = SU(2)_L \times SU(2)_R$  where spin connections
play the role of gauge fields and Riemann curvature tensors correspond to their field strengths,
it is possible to express the topological invariants in terms of $SU(2)_L$
and $SU(2)_R$ gauge fields \cite{opy,oh-yang,yang-corfu}
\begin{eqnarray} \label{euler-gauge}
\chi(M) &=& \frac{1}{4 \pi^2} \int_M  \Big( F^{(+)i}  \wedge
F^{(+)i}  - F^{(-)i}  \wedge F^{(-)i} \Big) + \frac{1}{4 \pi^2} \int_{\partial M}
\Big( a^{(+)i} - a^{(-)i} \Big) \wedge
\Big( F^{(+)i} + F^{(-)i} \Big) \nonumber \\
&& + \frac{1}{12 \pi^2} \int_{\partial M}
\varepsilon^{ijk} \Big( a^{(+)i} - a^{(-)i} \Big) \wedge
\Big( a^{(+)j} - a^{(-)j} \Big)
\wedge \Big( a^{(+)k} - a^{(-)k} \Big), \\
\label{signature-gauge}
\tau(M) &=& \frac{1}{6 \pi^2} \int_{M} \Big( F^{(+)i} \wedge F^{(+)i} +
F^{(-)i}  \wedge F^{(-)i} \Big) \nonumber \\
&& + \frac{1}{12 \pi^2} \int_{\partial M} \Big( a^{(+)i} - a^{(-)i} \Big) \wedge
\Big( F^{(+)i} - F^{(-)i} \Big) + \eta_S(\partial M),
\end{eqnarray}
where the fundamental 1-form (\ref{2-f-form}) is decomposed according to the Lie algebra splitting
$so(4) = su(2)_L \oplus su(2)_R$ as
\begin{equation}\label{f-form-dec}
v_{ab} \equiv a^{(+)i} \eta^i_{ab} + a^{(-)i} \overline{\eta}^i_{ab}
\end{equation}
and the volume forms are defined by $e^1 \wedge e^2 \wedge e^3 \wedge e^4 \equiv \sqrt{g} d^4 x$ and
$e^1 \wedge e^2 \wedge e^3|_{\partial M} \equiv \sqrt{h} d^3 x$.

We showed in Sec. III that $U(1)$ gauge fields for the NS and BN instantons can be written
in the form (\ref{u1f-gen}). Accordingly, given $U(1)$ gauge fields of the form (\ref{u1f-gen}),
one can calculate the gravitational metric that is given by Eq. (\ref{gmetric-gen})
or equivalently Eq. (\ref{metric-form}). Then, using the results of Appendix B, it is straightforward
to calculate the Euler density $\rho_\chi (M)$ in Eq. (\ref{euler-density}).
The results for the NS and BN instantons are, respectively, given by
\begin{eqnarray} \label{eden-ns}
\mathrm{NS}: && \rho_\chi(M) = - \frac{24 t^8}{(r^4 + t^4)^3}, \\
\label{eden-bn}
\mathrm{BN}: && \rho_\chi(M) = \frac{16t^8 (4r^4 + 8r^2 t^2 + 3t^4)}{r^2(r^2+ 2t^2)^3 (r^4 + r^2 t^2 + t^4)^2}
+ \frac{8t^8 (4r^4 + 7r^2 t^2 + 4t^4)^2}{(r^2+ 2t^2)^2 (r^4 + r^2 t^2 + t^4)^4}.
\end{eqnarray}
As expected, the Euler density for the NS instanton is regular everywhere while the Euler density for
the BN instanton is singular at the origin due to the first term in Eq. (\ref{eden-bn}).
But the remarkable fact is that the singular behavior of the Euler density is much milder, i.e. $1/r^2$,
than the Kretschmann scalar (\ref{bn-k}) which is $1/r^4$. Actually the $(1/r^2)$-singularity
in the Euler density (\ref{eden-bn}) is not harmful when we calculate the Euler characteristic (\ref{euler})
because the volume factor $\sqrt{g} d^4 x$ will safely cancel the singularity. Indeed we will get
a finite bulk contribution for the Euler characteristic $\chi(M)$ of the BN instanton. To be specific,
the bulk part of the Euler characteristic $\chi(M)$ is given by
\begin{equation}\label{euler-bulk}
    \chi_{bulk}(M) = \frac{1}{2\pi^2} \int_M \rho_\chi(M) e^1 \wedge e^2 \wedge e^3 \wedge e^4
\end{equation}
and we get the same value $\chi_{bulk}(M) = \frac{3}{2}$ for both the cases. Here we used the fact that
both the NS instanton and the BN instanton satisfy the ALE boundary condition $\mathbb{R}^4/\mathbb{Z}_2$
because they share the same asymptotic behaviors as was shown in Eqs. (\ref{asymp-ns}) and (\ref{asymp-bn})
and so $\int_{\mathbb{R}\mathbb{P}^3} \sigma^1 \wedge \sigma^2 \wedge \sigma^3 = \pi^2$.
Now let us look at the boundary terms in Eq. (\ref{euler}). The first boundary term will not contribute
because the curvature tensor will rapidly vanish $(\sim t^4/r^6)$ at infinity. But the second boundary term
will contribute and the explicit computation \cite{opy} gives the value
\begin{equation}\label{euler-bound}
    \chi_{boundary}(M) = \frac{1}{2\pi^2} \int_{\mathbb{R}\mathbb{P}^3}
    \sigma^1 \wedge \sigma^2 \wedge \sigma^3 = \frac{1}{2}.
\end{equation}
Note that the boundary contribution for the NS and BN instantons is also the same because they
satisfy the same asymptotic boundary condition.
In the end we get the same Euler characteristic (\ref{euler}) given by
\begin{equation}\label{euler-nsbn}
    \chi(M) = \frac{3}{2} + \frac{1}{2} = 2
\end{equation}
for both instantons.

The Hirzebruch signature $\tau(M)$ can be calculated similarly using
the result in Eq. (\ref{hirz-density}). But it is not necessary to
separately calculate the bulk part of the Hirzebruch signature
$\tau(M)$ for the NS instanton. One can easily check that the spin
connections in Eq. (\ref{spin-conn}) for the solution (\ref{esw-nsu1})
are anti-self-dual, i.e., $\omega_{ab} = - \frac{1}{2}
{\varepsilon_{ab}}^{cd} \omega_{cd}$ which automatically leads to
anti-self-dual curvature tensors. It should be the case because the
NS instanton is equivalent to the Eguchi-Hanson space which is an
ALE gravitational instanton. This means that $F^{(+)i} = 0, \;
\forall i$ in Eqs. (\ref{euler-gauge}) and (\ref{signature-gauge}) and so the relation
$\rho_\tau(M) = -  \frac{2}{3} \rho_\chi(M)$ is deduced. For the BN
instantons, one can directly calculate $\rho_\tau(M)$ in
Eq. (\ref{hirz-density}) using the result (\ref{bn-curvature}).
The result can be summarized as follows:
\begin{eqnarray} \label{hden-ns}
\mathrm{NS}: && \rho_\tau(M) = -  \frac{2}{3} \rho_\chi(M) = \frac{16 t^8}{(r^4 + t^4)^3}, \\
\label{hden-bn}
\mathrm{BN}: && \rho_\tau(M) = \frac{2t^8 (r^4 + 2r^2 t^2 + 2t^4)(4r^4 + 7r^2 t^2 + 4t^4)
(4r^6 + 9r^4 t^2 + 6r^2 t^4 + 2t^6)} {r^2(r^2+ 2t^2)^4 (r^4 + r^2
t^2 + t^4)^4} \nonumber \\
&& \qquad \qquad + \frac{2t^8 (4r^4 + 7r^2 t^2 + 4t^4)(4r^4 + 9r^2
t^2 + 4t^4)}{(r^2+ 2t^2)^2 (r^4 + r^2 t^2 + t^4)^4}.
\end{eqnarray}
As the Euler density (\ref{eden-bn}), the signature density
$\rho_\tau(M)$ in Eq. (\ref{hden-bn}) is also singular at the origin due to
the first term but it is not a harmful singularity either because we
will get a finite bulk contribution for the Hirzebruch signature
$\tau(M)$. To verify it,  consider the bulk part of the Hirzebruch
signature $\tau(M)$ given by
\begin{equation}\label{hirz-bulk}
    \tau_{bulk}(M) = \frac{1}{2\pi^2} \int_M \rho_\tau(M) e^1 \wedge e^2 \wedge e^3 \wedge e^4
\end{equation}
and we get the same value $\tau_{bulk}(M) = -1$ for both instantons.
The first boundary term in $\tau(M)$ vanishes for the same reason as
in the Euler characteristic $\chi(M)$. And the eta-invariant
$\eta_S(\partial M)$ is identically zero for the Eguchi-Hanson space
because $\eta_S(\partial M)$ for $k$ self-dual gravitational
instantons is given by \cite{gibb-perry}
\begin{equation}\label{eta-inv}
    \eta_S(\partial M) = - \frac{2 \epsilon}{3k} + \frac{(k-1)(k-2)}{3k}
\end{equation}
where $\epsilon =0$ for ALE boundary conditions and $\epsilon = 1$
for ALF boundary conditions. Although we did not explicitly
calculate the eta-invariant $\eta_S(\partial M)$ for the BN
instanton, it is reasonable to expect that it will also vanish
because the metric (\ref{bn-metric}) for the BN instanton shows
exactly the same asymptotic behavior as the Eguchi-Hanson space and
it also satisfies the ALE boundary condition. Thus we conclude that
\begin{equation}\label{hirz-nsbn}
    \tau(M) = -1 + 0 = -1
\end{equation}
for both NS and BN instantons.

Let us discuss some possible implications for the topological
invariants of $U(1)$ gauge fields. Our result (\ref{euler-nsbn}) for
the BN instanton strongly supports the topology change of spacetime
speculated by Braden and Nekrasov \cite{bn-inst}. Recall that the
Euler characteristic $\chi(M)$ can be determined by the set of nuts
and bolts through the fixed point theorem (see Eq. (4.6) in
Ref. \cite{bolt})
\begin{equation}\label{euler-fixed}
    \chi(M) = 2 \; \sharp(\mathrm{bolts}) + \sharp(\mathrm{nuts}).
\end{equation}
Therefore the result (\ref{euler-nsbn}) implies that the BN
instanton contains a noncontractible two-sphere $\mathbb{S}^2$
which is realized as a bolt in the gravitational solution
(\ref{bn-metric}) as we observed before. After such a blow-up of
$\mathbb{C}^2$ with $\mathbb{S}^2$, the resulting space becomes
K\"ahler and we showed before that the metric (\ref{bn-metric}) is
indeed K\"ahler. Thus the emergent gravity approach provides a more
accessible realization for the topology change of spacetime through $U(1)$
instantons.

It might be emphasized again that symplectic $U(1)$ gauge fields carry
exactly the same topological invariants as four-manifolds. But those
invariants are exotic from the gauge theory point of view because
they are represented by higher derivative terms of $U(1)$ field
strength $F_{\mu\nu}$. (Actually this issue was posed before in the
last of Sec. 5 in Ref. \cite{sty-06}.) Nevertheless, their properties are rather
natural when we recall that the ALE space is an orbifold resolution and the
Chern classes of orbifold bundles are typically rational numbers.
If one works with ``stacky" boundary conditions on the ALE space (by compactifying
with an orbifold line at infinity as in the work of Nakajima \cite{nakajima}),
then such properties actually arise and are in fact desired to match with corresponding
orbifold calculations.\footnote{We are indebted to an anonymous referee for this remark.}
Another interesting point is that there exist two independent topological invariants,
$\chi(M)$ and $\tau(M)$, for four-manifolds while the second Chern class $c_2(E)$ is a
unique topological invariant for the vector bundle $E$ of gauge fields.
Only for self-dual four-manifolds satisfying Eq. (\ref{g-inst}), two
invariants are related to each other. For example, closed half-flat manifolds satisfy
the relation $\chi(M) = \frac{3}{2} |\tau(M)|$ whereas noncompact half-flat manifolds obey
$\chi(M) = |\tau(M)| + 1$ \cite{opy,yang-corfu}. But, for general four-manifolds,
$\chi(M)$ and $\tau(M)$ are independent of each other. In addition, our explicit computation verifies that
it is necessary to include not only bulk terms  but also boundary contributions in order
to get integer-valued topological invariants.
All these features are unusual and interesting from the gauge theory perspective and so
further studies are required.

\section{Discussion}

The symplectic structure of spacetime is arguably the essence of
emergent gravity realizing the duality between general relativity and NC $U(1)$ gauge theory.
Our circumstantial test of emergent gravity picture reveals that the NC spacetime (\ref{nc-space})
must be taken seriously as a pillar for quantum gravity.
Moreover, we have to regard the NC algebra (\ref{nc-space}) as a raw precursor to the fabric of
spacetime that is coalesced into an organized form that we recognize as spacetime. Our analysis comparing
the NS instanton and the BN instanton indicates that a spacetime singularity in general relativity
can be resolved through the topology change of spacetime as far as a microscopic spacetime is NC.

From the viewpoint of emergent gravity, the topology of spacetime is determined by the topology
of $U(1)$ gauge fields on NC spacetime.
As is now well-known, the topology of NC $U(1)$ gauge fields is nontrivial and rich \cite{nc-top1,nc-top2}.
NC $U(1)$ instantons are the pith of the nontrivial topology of NC $U(1)$ gauge fields.
In this paper we observed that the nontrivial topology of NC $U(1)$ instantons
faithfully appears in the emergent gravity description. For example, NC $U(1)$ instantons
give rise to the ALE-type four-manifolds \cite{hsy-epl09,hsy-epjc09} whose nontrivial topology
is encoded in bolts (noncontractible two-cycles)
while NC $U(1)$ monopoles may be realized as the ALF-type four-manifolds
whose nontrivial topology is encoded in nuts (isolated points).
The nice formula (\ref{euler-fixed}) for the Euler characteristic clearly illuminates this aspect
of four-manifolds emergent from symplectic (or NC) $U(1)$ instantons or monopoles.

Then a natural question is about the status of spacetime singularity in NC spacetime. It is worthwhile
to notice that the NC space (\ref{nc-space}) is mathematically akin to the Heisenberg algebra $[x^i, p_j]
= i \hbar \delta^i_j$ in quantum mechanics where $\theta$ plays the role of $\hbar$. Thus one can expect
that the NC effect will be significant in a strongly gravitating regime,
typically near the spacetime singularities.
From the analogy between the NC spacetime and quantum mechanics, one can expect that there will be
a vital spacetime uncertainty relation as an analogue of the famous Heisenberg's uncertainty
relation $\Delta x \Delta p \geq \hbar$. This spacetime uncertainty relation gives rise to UV/IR mixing
in NC gauge theory \cite{uv/ir-mix} and is responsible for the holographic principle
in gravity \cite{holography1,holography2}. Therefore, in NC space, it is not possible to localize
a vast amount of energy to a point due to the spacetime exclusion.
The best way to realize a localized object in NC spacetime is to make a stable topological object
such as NC instantons \cite{ns-inst,nek-cmp03} or GMS solitons \cite{gms-sol}.
But the topological objects in NC spacetime are regular solutions without any singularity
and carry a nontrivial topology \cite{nc-top1,nc-top2}.
As a result, if spacetime geometry is emergent from NC gauge fields, the spacetime singularity
in general relativity may be a fake effect caused by our naive way of working
in a purely commutative space.

The relation in Eq. (\ref{nc-curved}) suggests that NC gauge fields can be interpreted as
the field variables defined in a locally inertial frame and their commutative description corresponds
to the field variables in a laboratory frame represented in terms of general curvilinear coordinates.
It presents a very beautiful picture about NC gauge fields.
Since this property holds even for the BN instanton which is neither self-dual
nor anti-self-dual, it is reasonable to suspect that the formula (\ref{nc-curved}) can be applied
to generic NC gauge fields. Now we will show that the identity (\ref{nc-curved}) is true
for general $U(1)$ gauge fields if the metric (\ref{eff-metric}) is symmetric,
i.e. $G_{\mu\nu} = G_{\nu\mu}$. Let us start with the SW map (\ref{sw-equiv-0}):
\begin{equation}\label{sw-equiv-form}
\frac{1}{4} \int d^4 y \widehat{F}_{\mu\nu}(y) \widehat{F}^{\mu\nu}(y) = \frac{1}{4} \int d^4 x \sqrt{G} G^{\mu\rho}G^{\nu\sigma} F_{\mu\nu} F_{\rho\sigma}
\end{equation}
which can be simply deduced from the SW maps (\ref{eswmap}) and (\ref{measure-sw}).
Using the one-form basis defined by Eq. (\ref{eff-vierbein}) and the definition $F_{ab}
\equiv E^\mu_a E^\nu_b F_{\mu\nu}$,
one can rewrite the right-hand side of Eq. (\ref{sw-equiv-form}) as
\begin{equation}\label{equiv-form}
\frac{1}{4} \int d^4 x \sqrt{G} G^{\mu\rho}G^{\nu\sigma} F_{\mu\nu} F_{\rho\sigma}
= \frac{1}{4} \int d^4 x \sqrt{G} F_{ab}(x)F^{ab}(x).
\end{equation}
Here we consider the coordinates $x^\mu=x^\mu(y)$ as the dynamical variables defined by Eq. (\ref{cov-cod}).
After using the formula (\ref{measure-sw}), the SW map (\ref{sw-equiv-form}) then reduces to the
following identity
\begin{equation}\label{curved-sw-equiv}
\frac{1}{4} \int d^4 y \widehat{F}_{\mu\nu}(y) \widehat{F}^{\mu\nu}(y) = \frac{1}{4} \int d^4 yF_{ab}(y)F^{ab}(y).
\end{equation}
Incidentally the coordinates $y^\mu$ correspond to geodesic normal coordinates in a Riemannian manifold
with the metric (\ref{eff-vierbein}). We note that the identity (\ref{curved-sw-equiv}) is nothing but
the SW equivalence between commutative and NC descriptions because it was originally
derived from the SW map (\ref{sw-equiv-form}). Since both sides are using the same coordinate system
and positive definite, we can deduce a local form from Eq. (\ref{curved-sw-equiv})
which is the identity (\ref{nc-curved}).

We remark that the emergent gravity formulated by using the SW map as in Sec. II cannot be applied
to a general Riemannian metric for the following reason. First of all, the effective metric defined by
Eq. (\ref{eff-metric}) is not symmetric in general. To have a symmetric Riemannian metric from symplectic
gauge fields, the condition (\ref{g-symm}) has to be obeyed. This condition is reduced to
the form (\ref{symg-f}) in a frame
with $\theta^{\mu\nu} = \frac{\theta}{2}\eta^3_{\mu\nu}$ (that can always be achieved by performing
an SO(4) rotation). Then some metric components in Eq. (\ref{eff-metric}) in this frame
identically vanish, e.g. $G_{12} = G_{34} = 0$. One can check that the Taub-NUT metric (\ref{taub-nut}),
for example, does not belong to such a class of metrics. Therefore we need a generalization
in order to extend the bottom-up approach of emergent gravity \cite{our-jhep12}
to a general class of metrics.

\section*{Acknowledgments}

This work was supported by the National Research Foundation
of Korean (NRF) grant funded by the Korea government (MEST) through
the Center for Quantum Spacetime (CQUeST) of Sogang University with
grant number 2005-0049409. The research of HSY was also supported by
Basic Science Research Program through the National Research
Foundation of Korea (NRF) funded by the Ministry of Education,
Science and Technology (2011-0010597).

\appendix

\section{'t Hooft symbols}

Since we heavily use several properties of the 't Hooft symbols,
we reproduce here the appendix A in Ref. \cite{loy} for reader's convenience.
The explicit components of the 't Hooft symbols $\eta^i_{\mu\nu}$ and ${\overline \eta}^i_{\mu\nu}$
for $i = 1,2,3$ are given by
\begin{eqnarray} \label{tHooft-symbol}
\begin{array}{l}
{\eta}^i_{\mu\nu} = {\varepsilon}^{i4\mu\nu} + \delta^{i\mu}\delta^{4\nu}
- \delta^{i\nu}\delta^{4\mu}, \\
{\overline \eta}^{i}_{\mu\nu} = {\varepsilon}^{i4\mu\nu} - \delta^{i\mu}\delta^{4\nu}
+ \delta^{i\nu}\delta^{4\mu}
\end{array}
\end{eqnarray}
with ${\varepsilon}^{1234} = 1$. They satisfy the following relations
\begin{eqnarray} \label{self-eta}
&& \eta^{(\pm)i}_{\mu\nu} = \pm \frac{1}{2} {\varepsilon_{\mu\nu}}^{\rho\sigma}
\eta^{(\pm)i}_{\rho\sigma}, \\
\label{proj-eta}
&& \eta^{(\pm)i}_{\mu\nu} \eta^{(\pm)i}_{\rho\sigma} =
\delta_{\mu\rho}\delta_{\nu\sigma}
-\delta_{\mu\sigma}\delta_{\nu\rho} \pm \varepsilon_{\mu\nu\rho\sigma}, \\
\label{self-eigen}
&& \varepsilon_{\mu\nu\rho\sigma} \eta^{(\pm)i}_{\sigma\lambda} = \mp (
\delta_{\lambda\rho} \eta^{(\pm)i}_{\mu\nu} + \delta_{\lambda\mu} \eta^{(\pm)i}_{\nu\rho} -
\delta_{\lambda\nu} \eta^{(\pm)i}_{\mu\rho} ), \\
\label{eta-etabar}
&& \eta^{(\pm)i}_{\mu\nu} \eta^{(\mp)j}_{\mu\nu}=0, \\
\label{eta^2}
&& \eta^{(\pm)i}_{\mu\rho}\eta^{(\pm)j}_{\nu\rho} =\delta^{ij}\delta_{\mu\nu} +
\varepsilon^{ijk}\eta^{(\pm)k}_{\mu\nu}, \\
\label{eta-ex}
&& \eta^{(\pm)i}_{\mu\rho}\eta^{(\mp)j}_{\nu\rho} =
\eta^{(\pm)i}_{\nu\rho}\eta^{(\mp)j}_{\mu\rho}, \\
\label{eta-o4-algebra}
&& \varepsilon^{ijk} \eta^{(\pm)j}_{\mu\nu} \eta^{(\pm)k}_{\rho\sigma} =
    \delta_{\mu\rho} \eta^{(\pm)i}_{\nu\sigma} - \delta_{\mu\sigma} \eta^{(\pm)i}_{\nu\rho}
    - \delta_{\nu\rho} \eta^{(\pm)i}_{\mu\sigma} + \delta_{\nu\sigma} \eta^{(\pm)i}_{\mu\rho},
\end{eqnarray}
where $\eta^{(+)i}_{\mu\nu} \equiv \eta^i_{\mu\nu}$ and $\eta^{(-)i}_{\mu\nu} \equiv {\overline
\eta}^i_{\mu\nu}$.

If we introduce two families of $4 \times 4$ matrices defined by
\begin{equation} \label{thooft-matrix}
[T^i_+]_{\mu\nu} \equiv \eta^i_{\mu\nu}, \qquad [T^i_-]_{\mu\nu} \equiv {\overline
\eta}^i_{\mu\nu},
\end{equation}
the matrices in Eq. (\ref{thooft-matrix}) provide two independent spin $s=\frac{3}{2}$ representations
of $su(2)$ Lie algebra. Explicitly, they are given by
\begin{eqnarray} \label{t+}
&& T^{1}_+ =\begin{pmatrix}
      0 & 0 & 0 & 1 \\
      0 & 0 & 1 & 0 \\
      0 & -1 & 0 & 0 \\
      -1 & 0 & 0 & 0 \\
             \end{pmatrix}, \;\;
  T^{2}_+ = \begin{pmatrix}
      0 & 0 & -1 & 0 \\
      0 & 0 & 0 & 1 \\
      1 & 0 & 0 & 0 \\
      0 & -1 & 0 & 0 \\
    \end{pmatrix}, \;\;
  T^{3}_+ = \begin{pmatrix}
      0 & 1 & 0 & 0 \\
      -1 & 0 & 0 & 0 \\
      0 & 0 & 0 & 1 \\
      0 & 0 & -1 & 0 \\
    \end{pmatrix}, \\
\label{t-}
&& T^{1}_- = \begin{pmatrix}
      0 & 0 & 0 & -1 \\
      0 & 0 & 1 & 0 \\
      0 & -1 & 0 & 0 \\
      1 & 0 & 0 & 0 \\
    \end{pmatrix}, \;\;
  T^{2}_- =  \begin{pmatrix}
      0 & 0 & -1 & 0 \\
      0 & 0 & 0 & -1 \\
      1 & 0 & 0 & 0 \\
      0 & 1 & 0 & 0 \\
    \end{pmatrix}, \;\;
  T^{3}_- =  \begin{pmatrix}
      0 & 1 & 0 & 0 \\
      -1 & 0 & 0 & 0 \\
      0 & 0 & 0 & -1 \\
      0 & 0 & 1 & 0 \\
    \end{pmatrix}
\end{eqnarray}
according to the definition (\ref{tHooft-symbol}).
The matrices in Eqs. (\ref{eta^2}) and (\ref{eta-ex}) immediately show that
$T^i_\pm$ satisfy $su(2)$ Lie algebras, i.e.,
\begin{equation} \label{thooft-su2}
[T^i_\pm, T^j_\pm] = - 2 \varepsilon^{ijk} T^k_\pm,
\qquad [T^i_\pm, T^j_\mp] = 0.
\end{equation}

\section{Spin connections and curvature tensors}

In this appendix, we calculate the spin connections and curvature tensors for the metric (\ref{metric-form}).
The spin connections are determined by solving the torsion free condition
\begin{equation}\label{torsion-free}
    T^a \equiv de^a + {\omega^a}_b \wedge e^b = 0.
\end{equation}
Explicitly they are given by
\begin{eqnarray}\label{spin-conn}
&& \omega_{14} = \frac{B' r + B}{A} \sigma^1, \qquad
\omega_{12} = \frac{A^2 - 2 B^2}{B^2} \sigma^3, \nonumber \\
&& \omega_{24} = \frac{B' r + B}{A} \sigma^2, \qquad \omega_{31} = - \frac{A}{B} \sigma^2, \\
&& \omega_{34} = \frac{A' r + A}{A} \sigma^3,  \qquad \omega_{23} = - \frac{A}{B} \sigma^1, \nonumber
\end{eqnarray}
where ${}^\prime \equiv \frac{d}{dr}$. The curvature tensor is then defined by
\begin{equation}\label{r-curvature}
    {R^a}_b = d{\omega^a}_b + {\omega^a}_c \wedge {\omega^c}_b.
\end{equation}
The explicit results are given by
\begin{eqnarray}\label{exp-curvature}
R_{12} &=& \frac{1}{r^2 B^2} \Bigl[ 4 - 3 \Bigl(\frac{A}{B} \Bigr)^2
      -  \Bigl( \frac{B' r + B}{A} \Bigr)^2 \Bigr] e^1 \wedge e^2
      + \frac{2}{rAB^3} (AB'-A'B) e^3  \wedge e^4, \nonumber \\
R_{31} &=& \frac{1}{r^2 B^2} \Bigl[ \Bigl( \frac{A}{B} \Bigr)^2
      - \frac{(A' r + A)(B' r + B)B}{A^3} \Bigr] e^3 \wedge e^1
      - \frac{1}{rAB^3} (AB'-A'B) e^2  \wedge e^4, \nonumber\\
R_{14} &=& \frac{rA'B'+ A'B - 2AB' - rAB''}{rA^3B}  e^1  \wedge e^4 + \frac{1}{r^2 B^2} \Bigl[
      \frac{A'r + A}{A} - \frac{B' r + B}{B} \Bigr] e^2 \wedge e^3, \qquad \\
R_{23} &=& \frac{1}{r^2 B^2} \Bigl[ \Bigl( \frac{A}{B} \Bigr)^2
      - \frac{(A' r + A)(B' r + B)B}{A^3} \Bigr] e^2 \wedge e^3
      - \frac{1}{rAB^3} (AB'-A'B) e^1  \wedge e^4, \nonumber \\
R_{24} &=& \frac{rA'B'+ A'B - 2AB' - rAB''}{rA^3B}  e^2  \wedge e^4 + \frac{1}{r^2 B^2} \Bigl[
      \frac{A'r + A}{A} - \frac{B' r + B}{B} \Bigr] e^3 \wedge e^1, \nonumber\\
R_{34} &=& - \frac{2}{r^2 B^2} \Bigl[ \frac{A'r + A}{A} - \frac{B' r + B}{B} \Bigr] e^1 \wedge e^2
+ \frac{1}{rA^4} \bigl(r(A')^2-rAA'' - AA' \bigr) e^3  \wedge e^4, \nonumber
\end{eqnarray}
where ${}'' \equiv \frac{d^2}{dr^2}$.

Using the above results, one can calculate the following quantities:
\begin{eqnarray} \label{euler-density}
    \rho_\chi (M) &\equiv & \frac{1}{64} \varepsilon^{abcd} \varepsilon^{efgh} R_{abef} R_{cdgh} \nonumber \\
 &=& \frac{1}{2} \bigl(R_{1234}^2 + R_{1423}^2 + R_{2431}^2 + R_{1212}R_{3434}
 + R_{3131}R_{2424}  + R_{1414}R_{2323} \bigr) \nonumber \\
  &=& \frac{1}{2r^4 A^6 B^6} \Bigl[ r B^2 \Bigl (3A^4 - 4A^2 B^2 +  B^2 \bigl(B + rB' \bigr)^2 \Bigl)
  \Bigl(- r(A')^2 + A \bigl( A' + r A'' \bigr) \Bigr) \nonumber \\
  && + 2 r B \Bigl( A^5 - B^3 \bigl(A + r A' \bigr) \bigl(B + r B' \bigr) \Bigr)
  \Bigl(-2 A B' + A'(B + r B' \bigr) - r A B'' \Bigr) \nonumber \\
 && + 6 r^2 A^4 \bigl( A B' - A' B \bigr)^2 \Bigr],
\end{eqnarray}
and
\begin{eqnarray} \label{hirz-density}
\rho_\tau (M) &\equiv & \frac{1}{48} \varepsilon^{cdef} R_{abcd} R_{abef} \nonumber \\
 &=& \frac{1}{3} \bigl(R_{1212}R_{1234} +  R_{1313} R_{2431} + R_{1414} R_{1423} + R_{2323} R_{2314}
  + R_{2424} R_{2431} + R_{3434} R_{1234} \bigr) \nonumber \\
  &=& \frac{2(AB'-A'B)}{3r^3 A^5 B^7} \Bigl[ r A^2 B^2 \Bigl (- r (B')^2 +  B \bigl(B' + rB'' \bigr) \Bigl)
  - 4 A^4 \bigl( A^2 - B^2 \bigr) \nonumber \\
  && + r B^4 \Bigl( r(A')^2 - A \bigl(A' + r A'' \bigr) \Bigr) \Bigr].
\end{eqnarray}


\begin{thebibliography}{99}


\bibitem{hsy-jhep09} H. S. Yang, J. High Energy Phys. {\bf 05} (2009) 012.

\bibitem{review4} H. S. Yang, Mod. Phys. Lett. A{\bf 25}, 2381 (2010).

\bibitem{hsy-jpcs12} H. S. Yang, J. Phys. Conf. Ser. {\bf 343}, 012132 (2012).

\bibitem{ncspace-ref} H. S. Snyder, Phys. Rev. {\bf 71}, 38 (1947);
C. N. Yang, {\it ibid.} {\bf 72}, 874 (1947).

\bibitem{dfr-cmp95} S. Doplicher, K. Fredenhagen and J. E. Roberts, Commun. Math. Phys.
{\bf 172}, 187 (1995).

\bibitem{b.greene} B. Greene, {\it The Elegant Universe: Superstrings, Hidden Dimensions,
and the Quest for the Ultimate Theory} (W.W. Norton \&  Company, Inc., 1999).

\bibitem{ncft1} M. R. Douglas and N. A. Nekrasov, Rev. Mod. Phys. {\bf 73}, 977 (2001).

\bibitem{ncft2} R. J. Szabo, Phys. Rep. {\bf 378}, 207 (2003).

\bibitem{string-book2} J. Polchinski, {\it String Theory Vol I \& II}
(Cambridge University Press, 1998).

\bibitem{sg-book1} V. I. Arnold, {\it Mathematical Methods of Classical
Mechanics} (Springer, New York, 1978).

\bibitem{sg-book2} R. Abraham and J. E. Marsden, {\it Foundations of Mechanics}
(Addison-Wesley, Reading, MA,  1978).

\bibitem{cornalba} L. Cornalba, Adv. Theor. Math. Phys. {\bf 4}, 271 (2000);
B. Jur\v co and P. Schupp, Eur. Phys. J. C{\bf 14}, 367 (2000).

\bibitem{hsy-ijmp09} H. S. Yang, Int. J. Mod. Phys. A{\bf 24}, 4473 (2009).

\bibitem{rivelles} V. O. Rivelles, Phys. Lett. B{\bf 558}, 191 (2003).

\bibitem{review1} R. J. Szabo, Class. Quantum Grav. {\bf 23}, R199 (2006).

\bibitem{review2} H. S. Yang, Mod. Phys. Lett. A{\bf 22}, 1119 (2007).

\bibitem{review3} H. S. Yang, Bulg. J. Phys. {\bf 35}, 323 (2008).

\bibitem{yasi-prd10} H. S. Yang and M. Sivakumar, Phys. Rev. D{\bf 82}, 045004 (2010).

\bibitem{lee-yang} J. Lee and H. S. Yang, {\it Quantum gravity from noncommutative spacetime},
[{\tt arXiv:1004.0745}].

\bibitem{review6} H. Steinacker, Class. Quantum Grav. {\bf 27}, 133001 (2010).

\bibitem{review7} J. Nishimura, Prog. Theor. Exp. Phys. {\bf 2012}, 01A101.

\bibitem{our-jhep12} S. Lee, R. Roychowdhury and H. S. Yang,
J. High Energy Phys. {\bf 09} (2012) 030.

\bibitem{eh-plb} T. Eguchi and A. J. Hanson, Phys. Lett. B{\bf 74}, 249 (1978).

\bibitem{eh-ap} T. Eguchi and A. J. Hanson, Ann. Phys. {\bf 120}, 82 (1979).

\bibitem{ns-inst} N. Nekrasov and A. Schwarz, Commun. Math. Phys.
{\bf 198}, 689 (1998).

\bibitem{sty-06} M. Salizzoni, A. Torrielli and H. S. Yang, Phys. Lett. B{\bf 634}, 427 (2006).

\bibitem{prl-06} H. S. Yang and M. Salizzoni, Phys. Rev. Lett. {\bf 96}, 201602 (2006).

\bibitem{bn-inst} H. W. Braden and N. A. Nekrasov, Commun. Math. Phys.
{\bf249}, 431 (2004).

\bibitem{ncft-sw} N. Seiberg and E. Witten, J. High Energy Phys. {\bf 09} (1999) 032.

\bibitem{t-symbol1} G. 't Hooft, Phys. Rev. D{\bf 14}, 3432 (1976).

\bibitem{t-symbol2} A. V. Belitsky, S. Vandoren and P. van Nieuwenhuizen,
Class. Quantum Grav. {\bf 17}, 3521 (2000).

\bibitem{loy} J. Lee, J. J. Oh and H. S. Yang, J. High Energy Phys. {\bf 12} (2011) 025.

\bibitem{dorn} O. Andreev and H. Dorn, Phys. Lett. B{\bf 476}, 402 (2000).

\bibitem{jsw01} B. Jur\v co, P. Schupp and J. Wess, Nucl. Phys. {\bf B604}, 148 (2001).

\bibitem{hsy-mpla06} H. S. Yang, Mod. Phys. Lett. A{\bf 21}, 2637 (2006).

\bibitem{ban-yan} R. Banerjee and H. S. Yang, Nucl. Phys. {\bf B708}, 434 (2005).

\bibitem{nc-gravity} P. Aschieri, C. Blohmann, M. Dimitrijevic, F. Meyer, P. Schupp and J. Wess,
Class. Quantum Grav. {\bf 22}, 3511 (2005); P. Aschieri, M. Dimitrijevic, F. Meyer and J. Wess,
{\it ibid.} {\bf 23}, 1883 (2006).

\bibitem{nc-seiberg} N. Seiberg, J. High Energy Phys. {\bf 09} (2000) 003.

\bibitem{nek-cmp03} N. A. Nekrasov, Commun. Math. Phys. {\bf 241}, 143 (2003).

\bibitem{furu-ptp} K. Furuuchi, Prog. Theor. Phys. {\bf 103}, 1043 (2000).

\bibitem{furu-ptps} K. Furuuchi, Prog. Theor. Phys. Suppl. {\bf 144}, 79 (2001).

\bibitem{kly-jkps02} K. Y. Kim, B.-H. Lee and H. S. Yang, J. Korean Phys. Soc. {\bf 41},
290 (2002) 290.

\bibitem{kly-plb01} K. Y. Kim, B.-H. Lee and H. S. Yang, Phys. Lett. B{\bf 523}, 357 (2001).

\bibitem{bps-paper} E. B. Bogomol'nyi, Sov. J. Nucl. Phys. {\bf 24}, 449 (1976); M. K. Prasad
and C. M. Sommerfield, Phys. Rev. Lett. {\bf 35}, 760 (1975).

\bibitem{hsy-epl09} H. S. Yang, Europhys. Lett. {\bf 88}, 31002 (2009).

\bibitem{egh-report} T. Eguchi, P. B. Gilkey and A. J. Hanson, Phys. Rep. {\bf 66}, 213 (1980).

\bibitem{besse} A. L. Besse, {\it Einstein Manifolds} (Springer-Verlag, Berlin, 1987).

\bibitem{lee-yi} K. Lee and P. Yi, Phys. Rev. D{\bf 61}, 125015 (2000).

\bibitem{future-paper} S. Lee, R. Roychowdhury and H. S. Yang,
{\it Emergent gravity from bottom-up approach}, to appear.

\bibitem{inst-review1} N. Dorey, T. J. Hollowood, V. V. Khoze and M. P. Mattis,
Phys. Rep. {\bf 371}, 231 (2002).

\bibitem{inst-review2} E. J. Weinberg and P. Yi, Phys. Rep. {\bf 438}, 65 (2007).

\bibitem{ks-inst02} P. Kraus and M. Shigemori, J. High Energy Phys. {\bf 06} (2002) 034.

\bibitem{our-topch} S. Lee, R. Roychowdhury and H. S. Yang, Phys. Rev. D{\bf 87}, 126002 (2013).

\bibitem{topch1} J. Madore and L. A. Saeger, Class. Quantum Grav. {\bf 15}, 811 (1998).

\bibitem{topch2} H. Shimada, Nucl. Phys. {\bf B685}, 297 (2004).

\bibitem{martinec} E. J. Martinec, Found. Phys. {\bf 43}, 156 (2013).

\bibitem{bolt} G. W. Gibbons and S. W. Hawking, Commun. Math. Phys. {\bf 66}, 291 (1979).

\bibitem{chu-npb02} C.-S. Chu, V. V. Khoze and G. Travaglini, Nucl. Phys. {\bf B621}, 101 (2002).

\bibitem{opy} J. J. Oh, C. Park and H. S. Yang, J. High Energy Phys. {\bf 04} (2011) 087.

\bibitem{oh-yang} J. J. Oh and H. S. Yang, Mod. Phys. Lett. A{\bf 28}, 1350097 (2013).

\bibitem{yang-corfu} H. S. Yang, PoS (CORFU2011) 063.

\bibitem{gibb-perry} G. W. Gibbons and M. J. Perry, Phys. Rev. D{\bf 22}, 313 (1980).

\bibitem{nakajima} H. Nakajima, Mosc. Math. J. {\bf 7}, 699 (2007).

\bibitem{nc-top1} J.A. Harvey, {\it Topology of the gauge group in noncommutative gauge theory},
[{\tt hep-th/0105242}].

\bibitem{nc-top2} F. Lizzi, R. J. Szabo and A. Zampini, J. High Engery Phys. {\bf 08} (2001) 032.

\bibitem{hsy-epjc09} H. S. Yang, Eur. Phys. J. C{\bf 64}, 445 (2009).

\bibitem{uv/ir-mix} S. Minwalla, M. Van Raamsdonk and N. Seiberg,
J. High Engery Phys. {\bf 02} (2000) 020.

\bibitem{holography1} G. 't Hooft, {\it Dimensional reduction in quantum gravity}, [{\tt gr-qc/9310026}].

\bibitem{holography2} L. Susskind, J. Math. Phys. {\bf 36}, 6377 (1995).

\bibitem{gms-sol} R. Gopakumar, S. Minwalla and A. Strominger, J. High Engery Phys. {\bf 05} (2000) 020.

\end{thebibliography}
\end{document}